\documentclass[]{emulateapj}

\usepackage{natbib}
\usepackage{bm}
\newcommand{\eqref}[1]{(\ref{#1})}

\shorttitle{}
\shortauthors{}

\begin{document}

\title{How AGN Jets Heat the Intracluster Medium -- Insights from Hydrodynamic Simulations}

\author{H.-Y.\ Karen Yang\altaffilmark{1,2},
and Christopher S.\ Reynolds\altaffilmark{2}} 
\altaffiltext{1}{Einstein Fellow}
\altaffiltext{2}{University of Maryland, College Park, Department of Astronomy and Joint Space Science Institute} 
\email{Email: hsyang@astro.umd.edu}

\begin{abstract}

Feedback from active galactic nuclei (AGN) is believed to prevent catastrophic cooling in galaxy clusters. However, how the feedback energy is transformed into heat, and how the AGN jets heat the intracluster medium (ICM) isotropically, still remain elusive. In this work, we gain insights into the relative importance of different heating mechanisms using three-dimensional hydrodynamic simulations including cold gas accretion and momentum-driven jet feedback, which are the most successful models to date in terms reproducing the properties of cool cores.
We find that there is net heating within two `jet cones' (within $\sim 30^\circ$ from the axis of jet precession) where the ICM gains entropy by shock heating and mixing with the hot thermal gas within bubbles. Outside the jet cones, the ambient gas is heated by weak shocks, but not enough to overcome radiative cooling, therefore forming a `reduced' cooling flow. Consequently, the cluster core is in a process of `gentle circulation' over billions of years. Within the jet cones, there is significant adiabatic cooling as the gas is uplifted by buoyantly rising bubbles; outside the cones, energy is supplied by inflow of already-heated gas from the jet cones as well as adiabatic compression as the gas moves toward the center. In other words, the fluid dynamics self-adjusts such that it compensates and transports the heat provided by the AGN, and hence no fine-tuning of the heating profile of any process is necessary. Throughout the cluster evolution, turbulent energy is only at the percent level compared to gas thermal energy, and thus turbulent heating is not the main source of heating in our simulation.

\end{abstract}



\section{Introduction}

Galaxy clusters with short central cooling times, or cool-core (CC) clusters, are expected to host massive gas inflows that are absent observationally, i.e., the 'cooling-flow problem' \citep{Fabian94}. The classical cooling-flow model predicts typical mass deposition rates of hundreds to a thousand solar masses per year. However, observationally there is a dearth of gas below $\sim 1$\ keV \citep[e.g.,][]{Peterson03}, and the star formation rates within observed CCs are suppressed by roughly an order of magnitude \citep{McNamara89, ODea08, McDonald11b, Hoffer12}. Various heating mechanisms have been proposed in order to circumvent catastrophic cooling \citep[][for a detailed review]{McNamara07}. However, processes that do not involve feedback \citep[e.g.,][]{Burns08}, such as thermal conduction \citep{Narayan01, Zakamska03, Voigt04}, often require fine-tuning \citep{Stewart84, Bregman88, Soker03} and likely only provide part of the heating if any.    


It is now widely accepted that feedback from active galactic nuclei (AGN) at cluster centers holds the key to solving the cooling flow problem. This is strongly motivated by the prevalence of AGN jet-inflated X-ray cavities/bubbles within CC clusters \citep{Best07, Mittal09, Birzan12} and the correlation between the cavity power (which measures the mechanical power of AGN jets) and the X-ray luminosity within the cores \citep{Rafferty06, Dunn08}. Theoretically, models that incorporate self-regulated energy injections from the central supermassive black holes (SMBHs) successfully prevent run-away cooling \citep{Sijacki07, Cattaneo07, Dubois10, Yang12b}. More recent simulations \citep{Gaspari12, Li14, Gaspari15b, Prasad15} that are based on SMBH accretion of cold gas, i.e., the `cold-mode' feedback \citep{Pizzolato05}, combined with momentum-driven AGN jets, further reproduce the positive temperature gradients \citep{Allen01b, Cavagnolo08, Hudson10}, the signatures of multiphase gas of the observed CCs \citep[e.g.,][]{Edge01, Salome03, Fabian06, McDonald11a}, and the slope of the soft X-ray spectrum \citep[e.g.,][]{Peterson03, Sanders10}.  Recent simulations by \cite{Yang15} that simultaneously include cooling, AGN feedback, and anisotropic conduction, show that even for unimpeded conductivity along magnetic field lines, conductive heating is likely significant only for the most massive clusters, while AGN are still the dominant sources of heating for most clusters. All the above highlight the importance of AGN feedback within CC clusters. 

Despite the successes, the details of how the AGN jets heat the intracluster medium (ICM) are still elusive. Many channels of AGN heating are proposed, including cavity heating \citep{Churazov01, Bruggen03}, shock heating \citep{Fabian03, Nulsen05, Randall15}, dissipation of sound waves \citep{Ruszkowski04b, Fabian05}, mixing by convection or turbulence \citep{David01, Kim03}, turbulent dissipation \citep{Dennis05, Zhuravleva14}, cosmic-rays \citep{Guo08, Pfrommer13}, and mixing with hot thermal gas within the bubbles \citep{Hillel16}. Some of the mechanisms only heat the ICM locally, e.g., cavity heating and bubble mixing should apply in the wakes and surrounding the bubbles, respectively. Moreover, it has been challenging for many processes to produce radial profiles of heating that match those of radiative cooling \citep{Brighenti02, Fabian05, Mathews06}. Therefore, it is most likely that multiple processes are involved to maintain the intricate balance between cooling and heating in the CCs.


However, two big questions remain: (1) What are the dominating heating mechanisms? (2) How do AGN jets heat the ICM {\it isotropically}, given that the jets are intrinsically directional \citep{Vernaleo06}? In this work, we attempt to address these questions by taking a closer look at one of the most successful models to date, namely, three-dimensional (3D) hydrodynamic simulations with cold-mode accretion and momentum-driven feedback. We utilize tracer particles and track specific terms in the hydrodynamic equations as guidance in order to better understand the heating and cooling processes as well as energy transport. We quantify the relative importance of all relevant processes, including radiative cooling, advection, adiabatic compression/expansion, shock heating, and mixing with hot bubble gas. We find that, though the AGN jets primarily heat regions surrounding the bubbles by mixing and weak shocks, the heat is isotropized by the process of `gentle circulation' within the cluster core.
Since such simulations have shown broad consistencies with observations, understanding how AGN heating works in the simulations may provide valuable insights into AGN heating in the real universe.   


The structure of the paper is as follows. We summarize the essential elements of our simulation and analyses in Section \ref{sec:method}. In Section \ref{sec:evolution}, we describe the overall cluster evolution and integrated properties. Then we present analyses from the Lagragian and Eulerian perspectives in Section \ref{sec:lagrangian} and \ref{sec:eulerian}, respectively. The relative contributions from relevant processes are quantified in Section \ref{sec:contribution}. We discuss the implications of our results on turbulent heating in Section \ref{sec:turb} and limitations of our simulation in Section \ref{sec:limitation}. Finally, we conclude our findings in Section \ref{sec:conclusion}.


\section{Methodology}
\label{sec:method}

We carry out a 3D hydrodynamic simulation including radiative cooling and self-regulated AGN feedback in an idealized Perseus-like cluster using the adaptive-mesh-refinement (AMR) code FLASH \citep{Flash, Dubey08}. The simulation setup is identical to the AGN-only simulation (Model A) in \cite{Yang15} (except that we neglect magnetic fields), to which we refer the readers for details. In the following we give a brief summary and emphasize methods that are specific to this study. 

The standard hydrodynamic equations are solved:
\begin{eqnarray}
&& \frac{\partial \rho}{\partial t} + \nabla \cdot (\rho {\bm v}) = 0,\\
&& \frac{\partial \rho {\bm v}}{\partial t} + \nabla \cdot \left( \rho {\bm v}{\bm v} \right) + \nabla P = \rho {\bm g},\\
&& \frac{\partial e}{\partial t} + \nabla \cdot \left[ (e+P){\bm v} \right] = \rho {\bm v} \cdot {\bm g} + {\mathcal H} - {\mathcal C},
\label{eq:hydro}
\end{eqnarray}
where ${\bm g}$ is the gravitational field, $\rho$, ${\bm v}$, $e=0.5\rho v^2 + e_{\rm i}$, and $P = (\gamma -1)e_{\rm i}$ are the density, velocity, total energy (including kinetic and internal energy), and thermal pressure of the gas. An equation of state for ideal gas with $\gamma=5/3$ is assumed. ${\mathcal H}$ and ${\mathcal C}$ are the heating and cooling terms, respectively. For radiative cooling, ${\mathcal C} = n_{\rm e}^2 \Lambda(T)$, where $n_{\rm e}=\rho/\mu_{\rm c}m_{\rm p}$ is the electron number density and $\Lambda(T)$ is the cooling function. In order to focus on hydrodynamic effects alone, magnetic fields and other transport effects (e.g., viscosity, conduction) are not included and will be considered in future work.    

The simulation domain is 1 Mpc on a side and is adaptively refined on steep temperature gradients up to an AMR refinement level of 7, which corresponds to a resolution element of 1.95 kpc. The diode boundary condition is used, which is similar to the outflow boundary condition but prohibits inflow of gas into the simulation domain. Radiative cooling is computed using the tabulated table of \cite{SutherlandDopita} assuming 1/3 solar metallicity. The gas profiles of the cluster are initialized using empirical fits to the observed Perseus cluster assuming hydrostatic equilibrium within a fixed NFW \citep{Navarro96} gravitational potential. The central SMBH is assumed to be fed by cold gas in its vicinity, i.e., the `cold-mode' accretion \citep[e.g.,][]{Gaspari12, Gaspari15, Li15, Voit15}. The cold gas ($T<5\times 10^5$\ K) is dropped out from the hot phase and replaced by passively-evolving tracer particles. The assumptions and consequences of this treatment are discussed in detail in \cite{Yang15}. The SMBH accretion rate is estimated by $\dot{M}_{\rm BH}=M_{\rm cold}/t_{\rm ff}$, where $M_{\rm cold}$ is the total amount of cold gas within an accretion radius of $r_{\rm accre}=4$\ kpc and $t_{\rm ff}=5$\ Myr is the approximate free-fall time at $r_{\rm accre}$ for Perseus. The accretion rate is then used to compute the mass, momentum, and energy returned by the bipolar AGN jets along the $z$-direction following these equations:
\begin{eqnarray}
\dot{M} &=& \eta \dot{M}_{\mathrm{BH}}, \nonumber \\
|\dot{P}| &=& \sqrt{2\eta \epsilon} \dot{M}_{\mathrm{BH}} c \label{eq:jets}\\
\dot{E} &=& \epsilon \dot{M}_{\mathrm{BH}} c^2, \nonumber
\label{eq:jet}
\end{eqnarray}
where $\eta$ is the mass loading factor and the injected energy is purely kinetic. We adopt $\eta=1$ and a feedback efficiency of $\epsilon=0.001$ for the current simulation. The feedback is applied to a cylinder with radius of 2.5 kpc and height of 4 kpc. A small angle ($15^\circ$) jet precession with a period of 10 Myr is assumed in order to improve coupling between the AGN feedback energy and the ICM. We note that the SMBH accretion and feedback prescriptions adopted here have been extensively investigated \citep[e.g.,][]{Gaspari12, Yang12, Li15, Prasad15, Yang15}. The chosen parameters have successfully reproduced the general properties of the CCs, such as self-regulation, positive temperature gradient, and the signatures of multiphase gas within the cluster cores. Therefore, the simulation is representative and suitable for studying how the heating and cooling balance each other in the self-regulated feedback process.  

We utilize passively evolving tracer particles in order to follow the thermodynamic history of the ICM. The tracer particles are initialized at the beginning of the simulations within a sphere of radius $r=100$\ kpc with particle spacings of 2 kpc. The particles record the hydrodynamic variables of the fluid along their trajectories. The particle data is dumped into output files at a frequency of 0.005 Gyr. 

In order to distinguish between AGN jets and the ICM, a tracer fluid is used to identify the AGN jets. For each grid cell, the mass fractions of the jets and the gas are denoted as $f_{\rm jet}$ and $f_{\rm gas}=1-f_{\rm jet}$, respectively. The value of $f_{\rm jet}$ is initially set to zero everywhere in the simulation domain. When the jets are launched, $f_{\rm jet}=1$ is set within the jet nozzle. With subsequent advection and mixing, regions influenced by the jets could then have values between 0 and 1. It follows that a mass-weighted quantity $X$ is defined as $\int f_{\rm gas} \rho X {\rm d}^3 x/ \int f_{\rm gas} \rho  \ {\rm d}^3 x$. 

One of the candidates for providing heat to the ICM is heating by shocks. As we will see in Section \ref{sec:evolution}, the shocks found in our simulation have typical Mach numbers below 2. We hereby briefly summarize how we estimate the ICM heating by weak shocks. We identify the shocks based on negative velocity divergence and jumps in gas pressure,
\begin{equation}
\frac{\Delta P}{P} = \frac{P_2-P_1}{P_1} = \frac{2\gamma}{\gamma+1} y,
\label{eq:dp}
\end{equation}
where $P_1$ and $P_2$ are the preshock and postshock pressure, respectively. The quantity $y$ is related to the Mach number by
\begin{equation}
y = \frac{\rho_1 v_{\rm s}^2}{\gamma P_1}-1 = \mathcal{M}^2-1,
\end{equation}
where $\rho_1$ is the preshock density and $v_{\rm s}$ is the shock speed. For reference, $\mathcal{M}=1.1, 1.5$ and 2.0 correspond to $\Delta P/P \sim 0.26, 1.56$ and 3.75, respectively. The jump in specific entropy for a weak shock is 
\begin{equation}
{\rm d} s \simeq \frac{2\gamma k_{\rm B}}{3(\gamma+1)^2 \mu m_{\rm H}} y^3,
\end{equation}
which then could be used to compute the amount of energy per unit volume gained by the gas, ${\rm d}H=\rho T {\rm d}s$. The volumetric heating rate by weak shocks is then estimated by ${\rm d}H/{\rm d}t$, where ${\rm d}t = 0.005$\ Gyr is the time interval between two subsequent output files. For the evaluation of shock heating (see Section \ref{sec:contribution}), we only include shocks with $\Delta P/P \ge 0.2$ as they account for most visible entropy jumps in the simulation (see Figure \ref{fig:mixing_shocks}). However, we verified that the amount of shock heating is insensitive to this choice because weak shocks around the threshold do not have a dominant contribution to heating. Note that some previous works apply additional criteria for diagnosing shocks \citep[e.g.,][]{Ryu03, Bruggen05, Bryan14} and therefore our estimates of shock heating should be considered as upper limits. 



\section{Results}
\label{sec:results}

\subsection{Overall evolution}
\label{sec:evolution}

We describe the overall evolution and integrated properties of the cluster in this section. Since other aspects of self-regulated AGN feedback in clusters have been extensively discussed in the literature \citep[e.g.,][]{Sijacki07, Gaspari11, Yang12b, Li14, Prasad15}, we focus on characteristics that are relevant for this study. 

\begin{figure}[tbp]
\begin{center}
\includegraphics[scale=0.55]{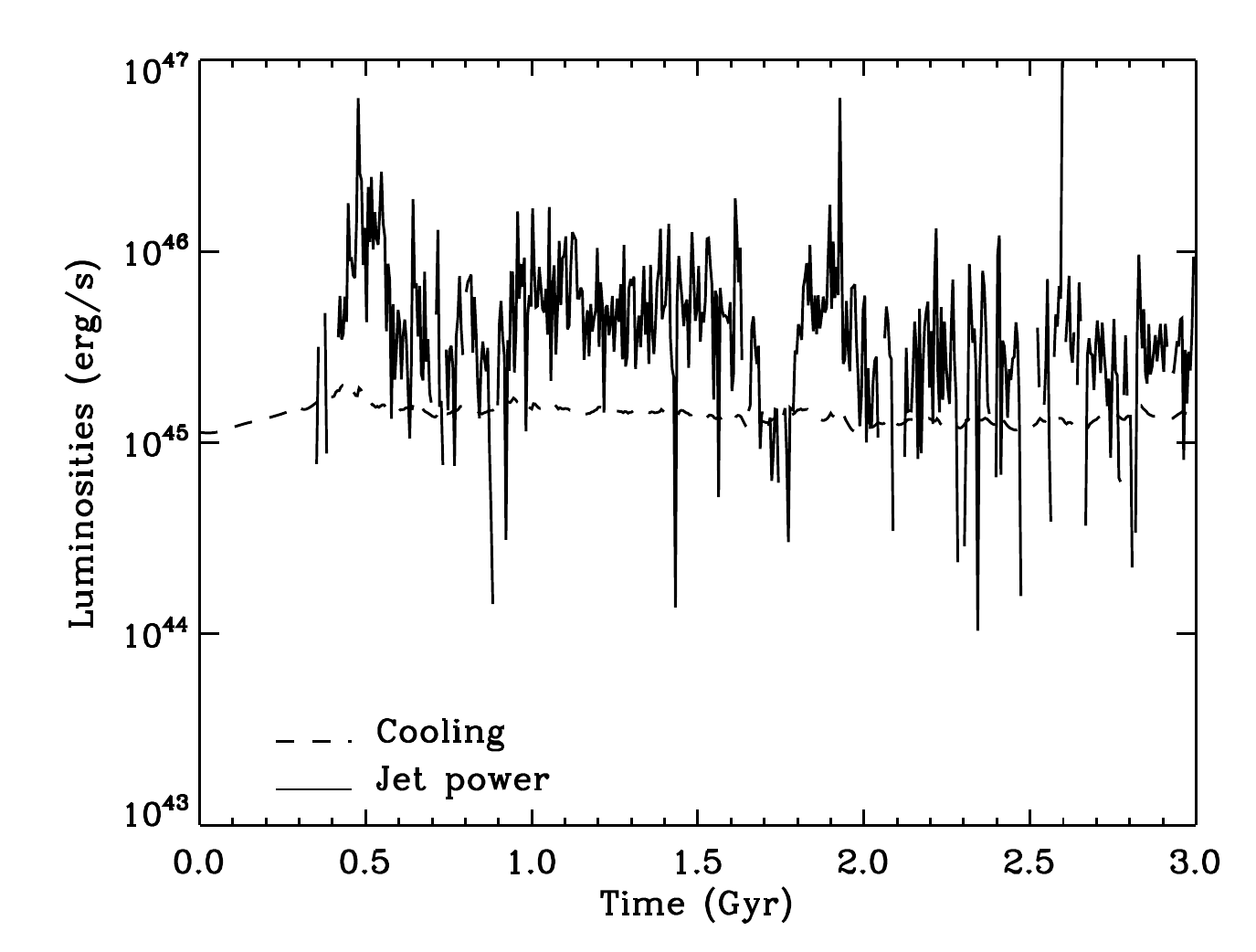} 
\caption{Evolution of the AGN jet power (solid line) and the X-ray luminosity integrated within 100\ kpc (dashed line). With self-regulated AGN feedback, the cluster reaches a quasi-equilibrium state after $t \sim 0.7$\ Gyr.}
\label{fig:pej}
\end{center}
\end{figure}

Figure \ref{fig:pej} shows the evolution of AGN jet power and X-ray luminosity within $r=100$\ kpc, which is about the cooling radius of the cluster (defined here as the radius at which the cooling time is 3\ Gyr). At the beginning of the simulation, the X-ray luminosity increases as the cluster contracts due to energy losses by radiative cooling, forming a cooling flow \citep{Fabian94, Li12}. The ICM cools and forms clumps of cold gas due to local thermal instabilities (TI) when $t_{\rm c}/t_{\rm ff} \lesssim 10$ \citep{McCourt12, Sharma12, Gaspari12, Li14, Meece15}. The central SMBH is fed by the cold gas in its vicinity and then subsequent jet activity is triggered at $t \sim 0.3$\ Gyr, which is approximately the initial cooling time of the cluster. The kinetic energy of the AGN jets is transformed into heat of the ICM (via processes that will be discussed in detail later), raising $t_{\rm c}/t_{\rm ff}$ above the threshold for TI and suppressing cold clump formation until the ICM cools again and fueling is resumed. Thereby a self-regulated AGN feedback cycle is established, maintaining the cluster in a quasi-equilibrium state after $t \sim 0.7$\ Gyr. \footnote{We note that, though our simulation setup is similar to \cite{Li14}, the evolution of the cluster is different owing to the different treatment of cold gas. Because in our simulation the dropped-out cold gas is efficiently uplifted by the rising bubbles and does not form massive disks surrounding the SMBH, quasi-equilibrium is more easily established, which is optimal for studying the long-term balance between heating and cooling.} After $t \sim 0.7$\ Gyr, the X-ray luminosity stays roughly constant ($\sim 1.3\times 10^{45}\ {\rm erg\ s^{-1}}$), and the average jet power is about $3 \times 10^{45}\ {\rm erg\ s^{-1}}$, implying $\sim 40\%$ of the kinetic energy is transformed into thermal energy within 100\ kpc. The average mass accretion rate for the simulated cluster is $\sim 15\%$ of the inferred mass deposition rate of the observed Perseus cluster \citep[][]{Allen01}, 
much suppressed compared to that of a pure cooling flow and consistent with the range of mass deposition rates \citep[e.g.,][]{Hudson10} and star formation efficiencies \citep[e.g.,][]{McDonald11b, Hoffer12} for observed CC clusters. 

\begin{figure*}[tbp]
\begin{center}
\includegraphics[scale=0.4]{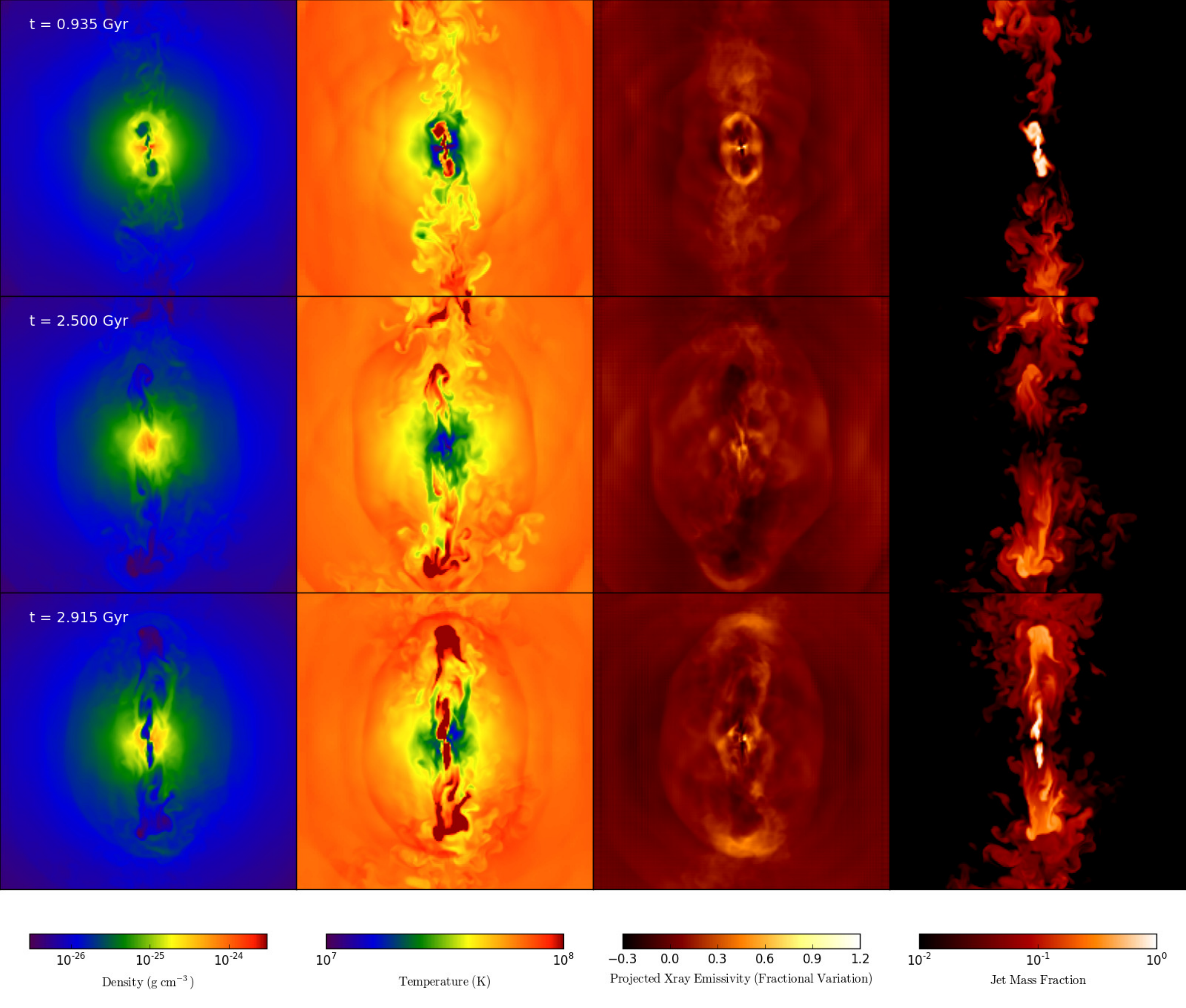} 
\caption{Columns from left to right are slices of gas density, temperature, projected X-ray emissivity (fractional variation from a radially averaged projected emissivity profile), and the jet mass fraction through the cluster center. Simulation snapshots at $t=0.935$, $t=2.500$, and $t=2.915$\ Gyr are shown in the top to bottom rows. Slices are 400\ kpc on a side. Lines in the last panel show the boundary of the jet cones, which are defined to be the region within $30^\circ$ away from the $z$-axis.}
\label{fig:slices}
\end{center}
\end{figure*}

Figure \ref{fig:slices} shows slices of gas density, temperature, projected X-ray emissivity, and the jet mass fraction ($f_{\rm jet}$). Snapshots are taken at three epochs, $t=0.935$, 2.500, and 2.915, which represent three characteristic phases of AGN activity: single event, quiescent, and multiple episodes. In general, the AGN jets inflate underdense bubbles, producing X-ray cavities similar to those observed \citep{Fabian00}. The cavities are filled with high fractions of jet materials that are hot ($\sim$ few tens of keV) due to the initial thermalization of the jet kinetic energy by shocks. When the bubbles rise and expand due to buoyancy, they mix with the ICM (i.e., $f_{\rm jet}$ decreases) as they are shredded due to Rayleigh-Taylor, Kelvin-Helmholtz, and Richtmyer-Meshkov instabilities. Weak shocks can also be seen in the temperature and X-ray maps. The Mach numbers of the weak shocks ($1 \lesssim \mathcal{M} < 2.1$) are consistent with those observed in nearby CC clusters \citep[e.g.,][]{Fabian06, Blanton09, Randall15}. 

Since our jets are directional by construction, there is a clear distinction between the ICM properties within and outside the `jet cones', which are defined as the region within $\sim 30^\circ$ away from the $z$-axis (see last panel in Figure \ref{fig:slices}). The buoyantly rising cavities containing partially mixed jet materials primarily lie within the jet cones; this is also where turbulence is generated by hydrodynamic instabilities during the disruption of bubbles \citep{Yang15}. Much of the ambient ICM (i.e., outside the jet cones) does not directly contact the jet materials, but is influenced by the weak shocks and sound waves. As we will show later, the primary heating mechanisms are very different between the jet cones and the ambient region. In reality, the bubbles may have a more isotropic distribution like those observed in Perseus \citep{Fabian00}. We will discuss the implications of our results in this aspect in Section \ref{sec:limitation}.

\begin{figure*}[tbp]
\begin{center}
\includegraphics[scale=0.55]{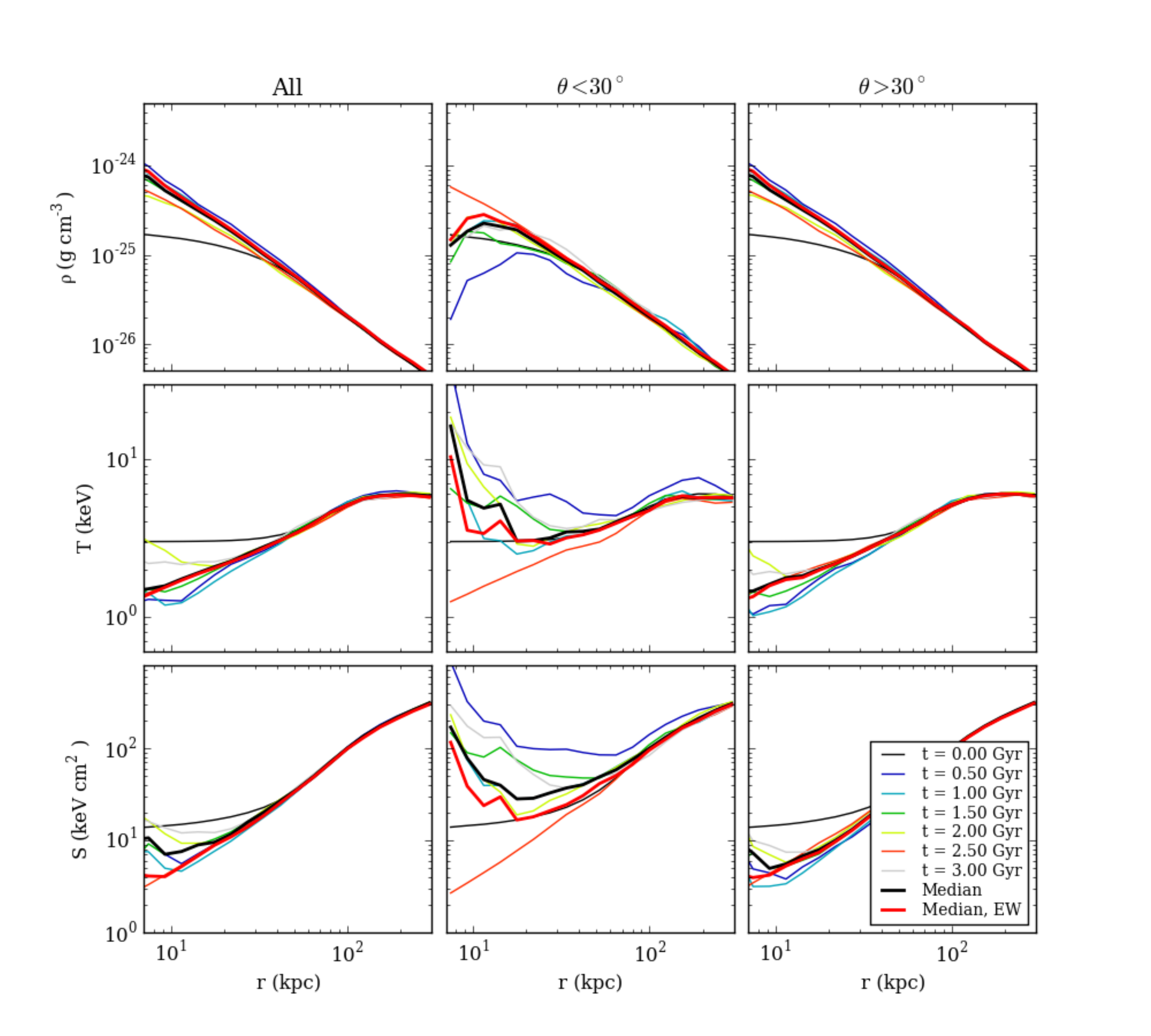} 
\caption{From top to bottom rows show the evolution of density, temperature, and entropy profiles. Left to right columns show profiles averaged over the entire cluster, the jet cones ($<30^\circ$ away from the $z$-axis; see last panel in Figure \ref{fig:slices}), and the ambient region (outside the jet cones), respectively. Thick lines represent the median of all profiles. All quantities are mass-weighted except for the red curves, which show the median values of the emission-weighted profiles. }
\label{fig:prf}
\end{center}
\end{figure*}

Figure \ref{fig:prf} shows the profiles of gas density, temperature, and entropy (top to bottom rows) averaged within the entire cluster (left column), the jet cones (middle column), and the ambient region (right column). The thin curves show mass-weighted profiles at different epochs. The black and red thick lines represent the median of the mass-weighted and emission-weighted profiles, respectively. The profiles integrated over the whole cluster are in agreement with observed CC clusters: the positive gradient of the temperature profiles is preserved, and the entropy profiles follow a baseline power-law with a floor \citep[e.g.,][but see also \cite{Panagoulia14}]{Cavagnolo09, Lakhchaura16}. The profiles (both mass-weighted and emission-weighted) look very similar for the entire cluster and the ambient region, implying the mass and emission are dominated by the denser ambient ICM. The influence of the AGN is more prominent within the jet cones, as expected. When the AGN is active (e.g., $t \sim 1$ and 3 Gyr; see top and bottom rows in Figure \ref{fig:slices}), the gas density is suppressed, and the temperature and entropy are raised. During the quiescent state (e.g., $t \sim 2.5$\ Gyr; middle row of Figure \ref{fig:slices}), the profiles are close to the baseline. 

\begin{figure}[tbp]
\begin{center}
\includegraphics[scale=0.2]{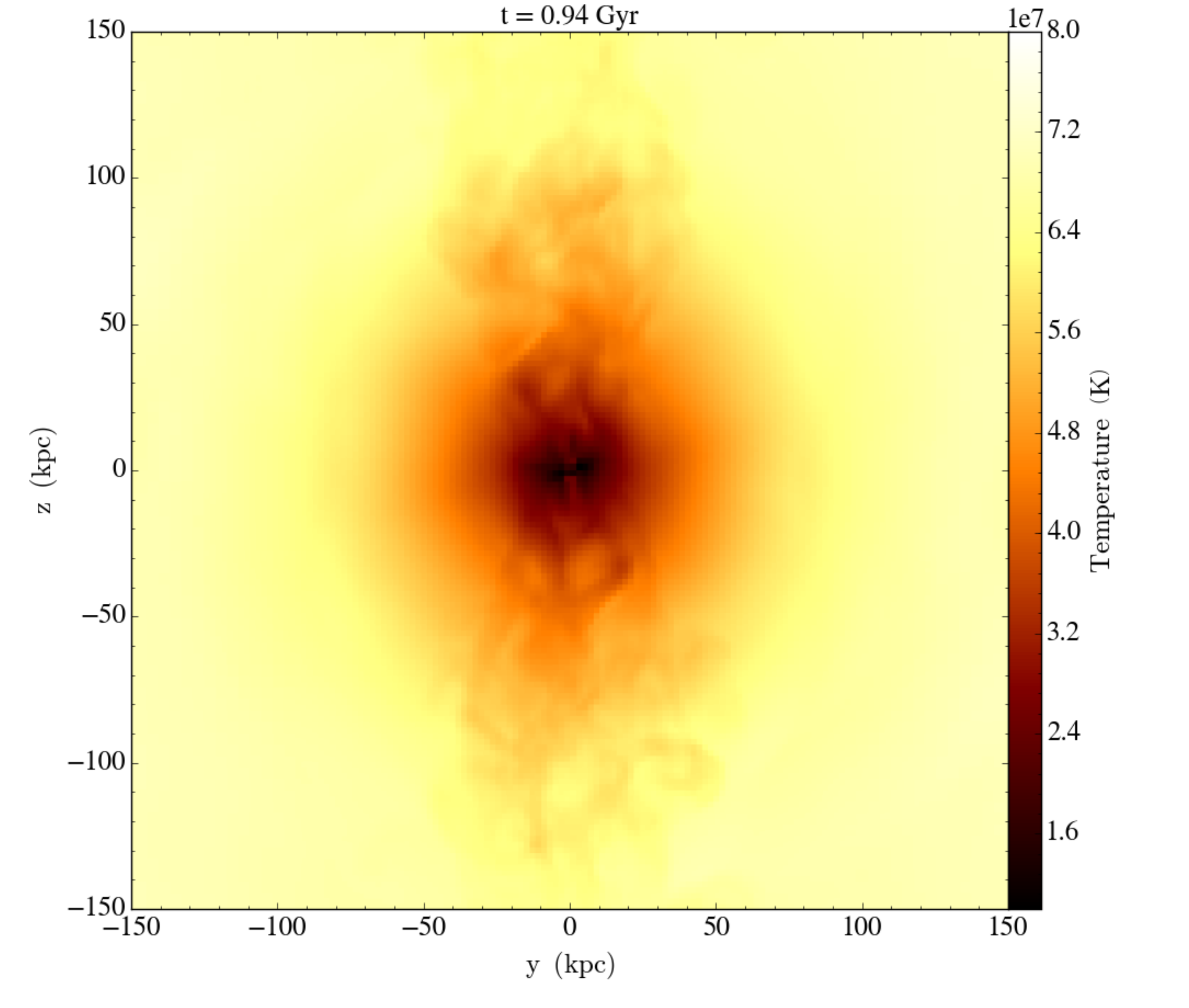} 
\caption{A zoomed-in image of projected, emission-weighted temperature map at $t=0.935$\ Gyr (to be compared with the second panel in the top row in Figure \ref{fig:slices}). }
\label{fig:Tmap}
\end{center}
\end{figure}

Note that even in the jet cones where most of the heating appears to occur, the emission-weighted profiles do not show significant temperature and entropy inversions (except the very inner $\sim 10$\ kpc where the effects of the jets at this radius are likely overestimated due to limited resolution). That is because the gas with very high temperature (see second columns in Figure \ref{fig:slices}) is mainly contained within the underdense bubbles, and is thus subdominant in terms of mass. The $n_{\rm e}^2$ dependence of the X-ray emissivity further attenuates the contribution from the hot gas in terms of emission. Line-of-sight projections also act to bury the signature of the hot gas. Figure \ref{fig:Tmap} shows a projected, emission-weighted temperature map of the same image in the top, second panel in Figure \ref{fig:slices}. As can be clearly seen, the projected location of the bubbles only have slightly higher temperature than their surroundings, instead of a significant trace of heating as one would naively expected. This is in good agreement with observed temperature distributions of well-studied cavities \citep{Fabian06, Blanton09}.  

We have demonstrated that kinetic AGN feedback indeed could self-regulate and yield gross ICM  properties that are consistent with observations. The quasi-equlibrium state of the cluster over Gyr timescales imply a rough balance between radiative cooling and AGN heating within the cluster core {\it as a whole}. In the following sections, we will look into details how this balance is established. 


\subsection{The Lagrangian view}
\label{sec:lagrangian}

In this section, we show results from analyzing the thermodynamic properties of the ICM tracer particles. In essence, these tracers evolve following the hydrodynamic equations in the Lagrangian form. Particularly, the evolution of entropy of the tracers directly probe the balance between heating and cooling, i.e., 
\begin{equation}
\rho T \frac{{\rm d} s}{{\rm d} t} = \mathcal{H}(l) - \mathcal{C}(l),
\end{equation}
where $s$ is the entropy, and $\mathcal{H}(l)$ and $\mathcal{C}(l)$ are the heating and cooling terms along trajectories of the tracer particles. The entropy is important because it tells us whether and where new sources of energy are {\it generated} (as opposed to being {\it transported}, see the next section). We will show in the following that there is net heating within the jet cones, while cooling dominates over heating for the ambient region. Also, the primary heating mechanisms within the jet cones are shocks as well as mixing between the ICM and the jet materials. 

\begin{figure*}[tbp]
\begin{center}
\includegraphics[scale=0.55]{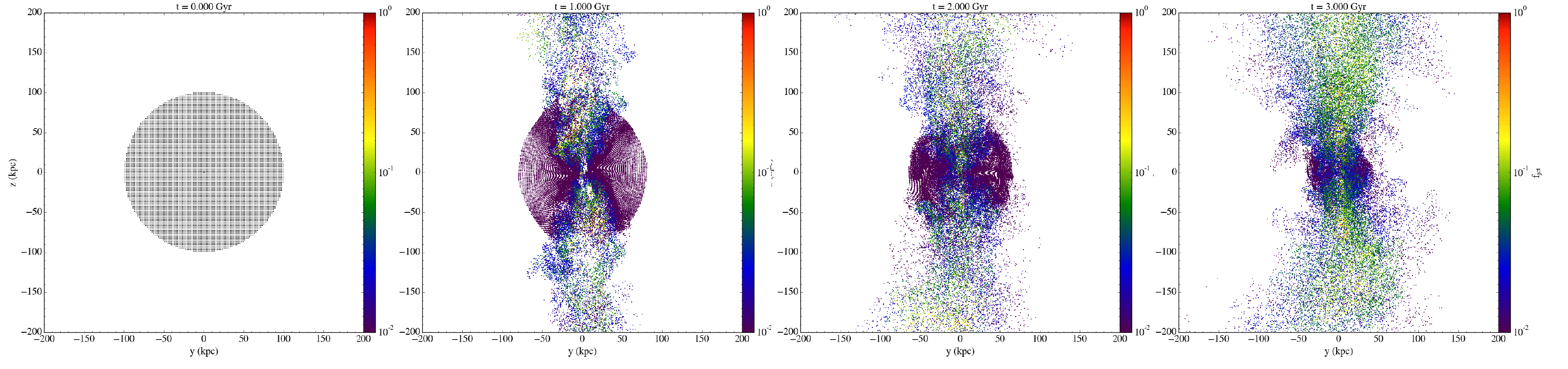} 
\caption{Evolution of the projected locations of the ICM tracer particles. Only particles with initial positions within the region [$r<100$\ kpc, -10\ kpc $\le x \le 10$\ kpc] are plotted for clarity. Colors represent the jet mass fraction ($f_{\rm jet}$) traced by the particles.}
\label{fig:tracers}
\end{center}
\end{figure*}

Figure \ref{fig:tracers} shows the evolution of the projected locations of the ICM tracers. The colors represent the jet mass fraction, $f_{\rm jet}$, at the locations of the particles. At $t=0$, the particles are uniformly distributed within 100 kpc and have $f_{\rm jet}=0$. Before the AGN is triggered at $t\sim 0.3$\ Gyr, the cluster contracts due to radiative cooling and thus the sphere containing the particles shrinks radially. Later on, the AGN jets break the symmetry. Particles tracing the ambient region (outside the jet cones, or having low $f_{\rm jet}$) keep contracting, similar to a cooling flow. However, note that the initial cooling time is 3 Gyr at $r=100$\ kpc. Therefore, the fact that the sphere bounding the ambient-gas particle tracers has a radius of $r \sim 40$\ kpc at $t=3$\ Gyr instead of sinking to the cluster center implies that the gas inflow is suppressed compared to a pure cooling flow. As we will show below (and also in Section \ref{sec:contribution}), though cooling is dominant within the ambient region, the cooling rate is partially compensated by heating from weak shocks.    

On the other hand, tracer particles that start from the jet cones or from smaller radii in the ambient region (those have enough time to reach the center by 3 Gyr), are uplifted to larger distances by the rising bubbles. As can be seen from Figure \ref{fig:tracers}, the particle distributions near the jet cones become broader over time. This is because of the combination of a few effects. First, particles that were lifted can slip toward the side of the bubbles and flow inward radially to fill the vacuum behind the cavities. As they flow in, they can be pushed sideways by elliptical shocks generated by subsequent AGN activity. Turbulent diffusion also helps to spread the particle distribution. Note that some of these particles eventually leave the jet cones and become parts of the ambient medium. If the simulation were run longer, they would cool to the center and be lifted by the bubbles again. In other words, there is a gentle {\it circulation} within the whole cluster. In Section \ref{sec:contribution}, we will show that this circulation plays a crucial role for the transport of heat.     

\begin{figure*}[ptb]
\begin{center}
\includegraphics[scale=0.8]{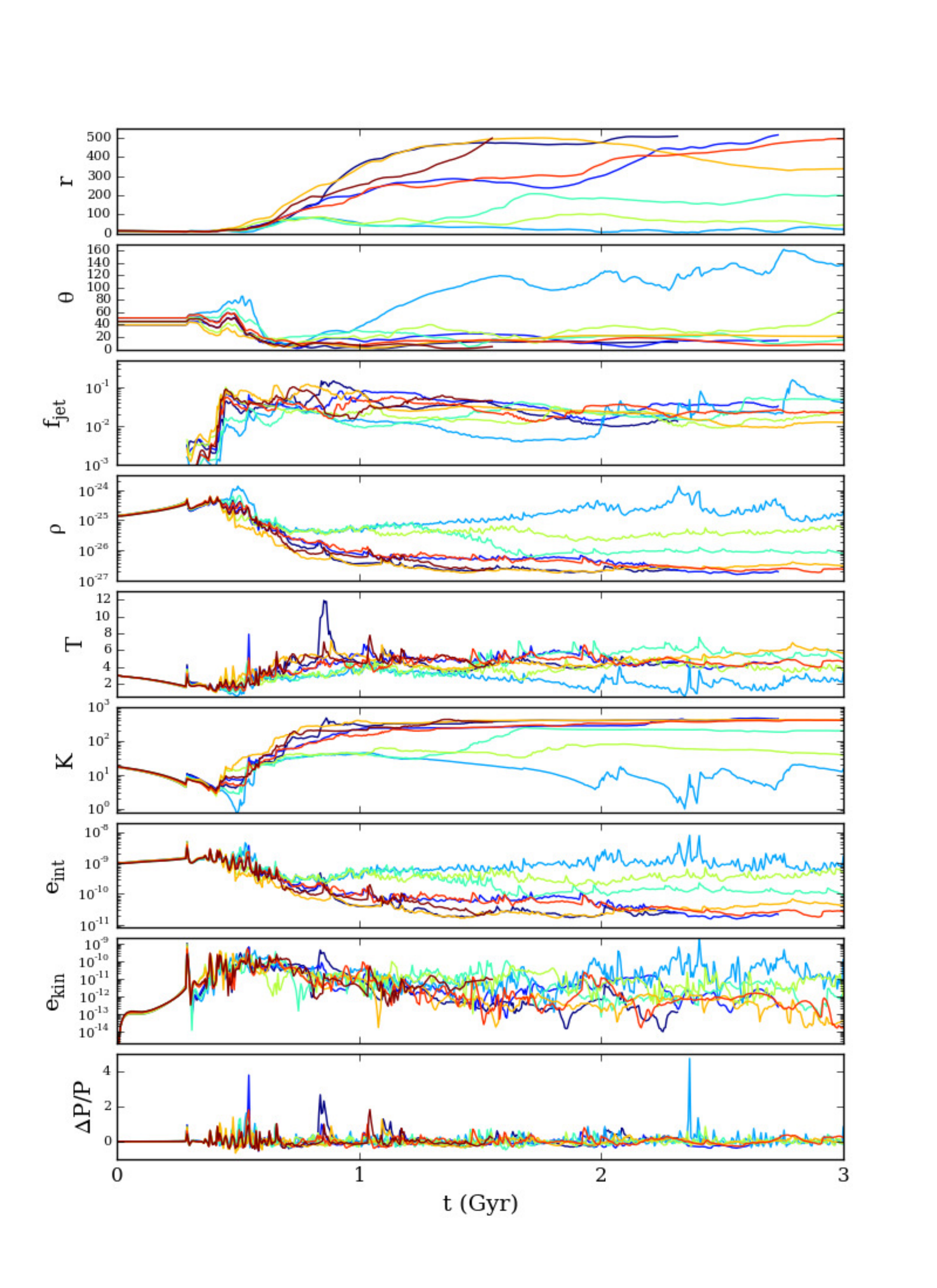} 
\caption{Evolution of all relevant variables of eight ICM tracers that are initially located $\sim 10$\ kpc from the cluster center right above the equatorial plane (TR10; see text for definition). From top to bottom the panels show the particle coordinates (the $r$ and $\theta$ components), the jet mass fraction ($f_{\rm jet}$), gas density ($\rho$), temperature ($T$), entropy ($K\equiv T/n_{\rm e}^{2/3}$), internal energy ($e_{\rm int}$), kinetic energy ($e_{\rm kin}$), and the pressure contrast ($\Delta P/P$) as a diagnostic for shocks (Equation \ref{eq:dp}).}
\label{fig:TR10}
\end{center}
\end{figure*}

\begin{figure}[ptb]
\begin{center}
\includegraphics[scale=0.55]{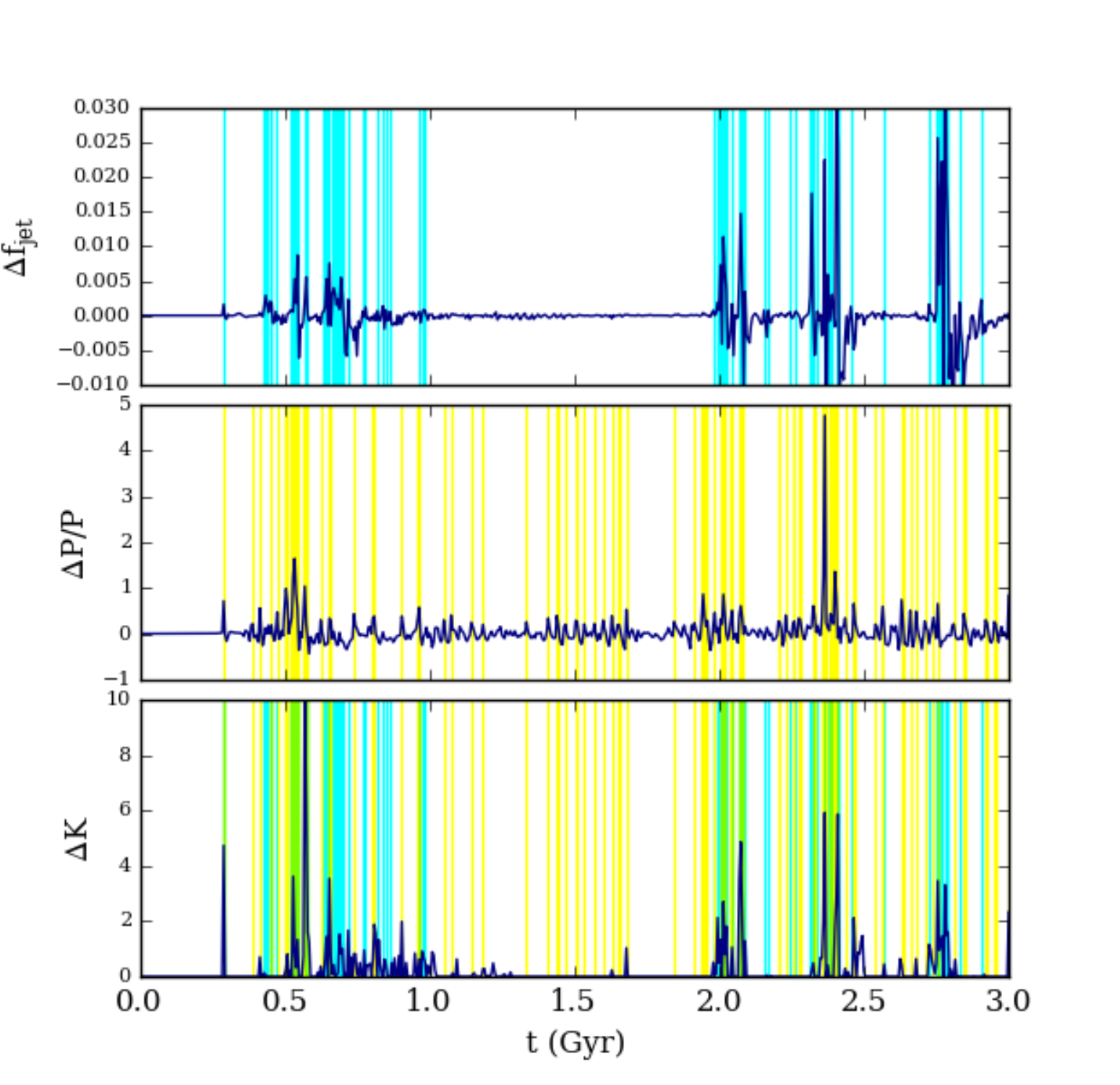} 
\caption{Evolution of one of the TR10 tracers (light blue curve in Figure \ref{fig:TR10}). The panels from top to bottom show jumps in $f_{\rm jet}$ (which traces mixing between the ICM and the jet materials), $\Delta P/P$ (which diagnoses shocks) and jumps in entropy, respectively. Cyan and yellow vertical lines mark the time when there is more significant mixing and shock heating, respectively. Both mixing and shocks are important sources of ICM heating within the jet cones.}
\label{fig:mixing_shocks}
\end{center}
\end{figure}

Next, we study in more detail the thermodynamic histories of the ICM tracers. Figure \ref{fig:TR10} shows the evolution of all relevant hydrodynamic variables along the trajectories of eight particles that have initial positions within the region [-2\ kpc $\le x \le 2$\ kpc, 8\ kpc $\le y \le 12$\ kpc, 8\ kpc $\le z \le 12$\ kpc]. Hereafter we denote them as TR10 since these particles are initially located $\sim 10$\ kpc from the center right above the equatorial plane. Consistent with the picture illustrated by Figure \ref{fig:tracers}, for the first $t\sim 0.45$\ Gyr before they mix with the jet materials, TR10 cool and flow in because {\it cooling dominates over heating within the ambient region} (i.e., the entropy $K\equiv T/n_{\rm e}^{2/3}$ decreases). A few weak shocks (i.e., jumps in $\rho, T, K, e_{\rm int}, e_{\rm kin}$, and $\Delta P/P$) heat the gas and raise the entropy, which slows down but does not completely stop the cooling flow. After TR10 start to enter the jet cones ($f_{\rm jet} \gtrsim 0.01$), the evolution becomes more complicated because several processes operate simultaneously. First, the entropy of the traced gas increases whenever $f_{\rm jet}$ increases (as can be seen clearly from the light blue curve), providing evidence for heating from mixing between the ICM and the jet materials. Second, weak shocks generated by repetitive AGN outbursts also add entropy to the gas. Though each shock only increases the entropy slightly, the shocks occur very frequently (see also Figure \ref{fig:mixing_shocks}). Lastly, as the gas is lifted away from the center by buoyantly rising bubbles, its density and internal energies decrease due to adiabatic expansion. All the above effects combined result in the evolution of the temperature as seen in Figure \ref{fig:TR10}. Note that as long as the particles stay within the jet cones (all but the light blue curve during $t\sim 1-2$\ Gyr), their entropy increases, meaning that on average {\it heating dominates over cooling within the jet cones}.  

\begin{figure*}[ptb]
\begin{center}
\includegraphics[scale=0.8]{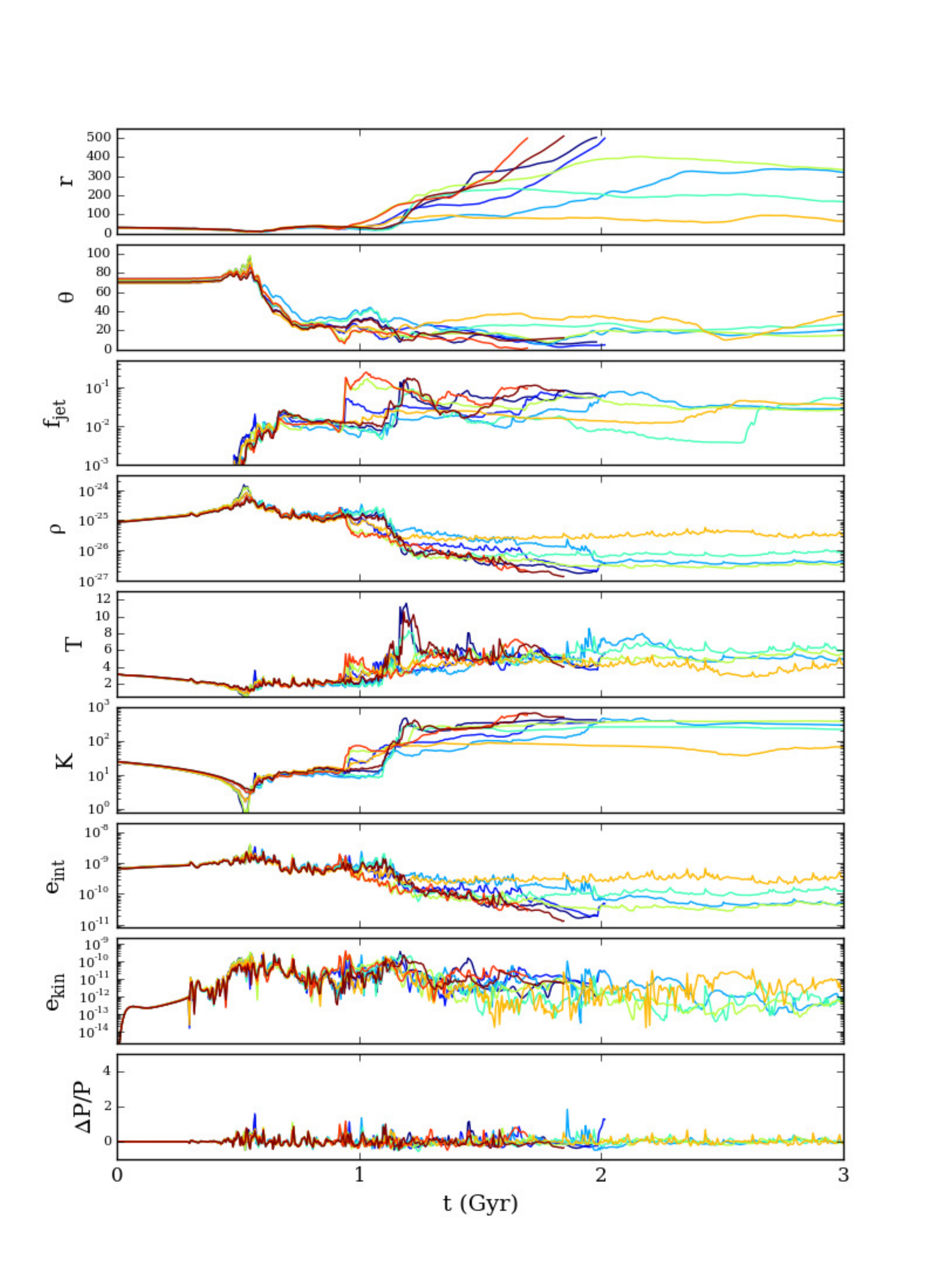} 
\caption{Same as Figure \ref{fig:TR10} but for ICM tracers with initial positions $\sim 30$\ kpc from the cluster center (TR30).}
\label{fig:TR30}
\end{center}
\end{figure*}

Figure \ref{fig:TR30} shows the same plot for tracers TR30 that are initially located within the region [-2\ kpc $\le x \le 2$\ kpc, 28\ kpc $\le y \le 32$\ kpc, 8\ kpc $\le z \le 12$\ kpc]. The general trends are similar to TR10, except that they start to mix with the jet fluid at a later time, $t\gtrsim 0.5$\ Gyr. Again we find that cooling dominates over heating during the initial period when TR30 are in the ambient region. After they enter the jet cones where heating prevails over cooling, their entropy increases with time almost monotonically. The heating within the jet cones again are done by mixing and shocks. We see significant entropy jumps associated with jumps in $f_{\rm jet}$ (e.g., red curve at $t \gtrsim 0.9$\ Gyr, brown and navy lines at $t \sim 1.15$\ Gyr). The shocks are weaker ($\Delta P/P < 1.5$ or $\mathcal{M} \lesssim 1.5$) but also contribute to heating (e.g., red curve at $t \sim 0.7-0.8$\ Gyr where $f_{\rm jet}$ stays roughly constant but there are small jumps in entropy coincident with shocks; see also Figure \ref{fig:mixing_shocks}). 

Note that the kinetic energy throughout the cluster evolution remains only at the percent level compared to the internal energy (see Figure \ref{fig:TR10} and \ref{fig:TR30}). This is consistent with a recent work by \cite{Reynolds15} that have found that driving of turbulence by AGN in the ambient gas is inefficient. Even if we assume all the kinetic energy goes into turbulent heating (in reality the kinetic energy is also contributed by bulk motions), it is not enough to provide substantial heat to the cluster gas. Therefore, turbulent heating is not the primary source of ICM heating in our simulation. We will discuss this result in more detail in Section \ref{sec:turb}. 

We now have a closer look at heating from mixing and shocks within the jet cones, as some of the entropy jumps may not be apparent in Figure \ref{fig:TR10} and \ref{fig:TR30}. In order to see clear associations among entropy jumps, mixing and shocks, we plotted in Figure \ref{fig:mixing_shocks} the difference in $f_{\rm jet}$, $\Delta P/P$ and the difference in entropy for one of TR10, namely, the light blue curve in Figure \ref{fig:TR10}. The difference is done between two consecutive output files (i.e., every $\Delta t=0.005$\ Gyr). Jumps in $f_{\rm jet}$ greater than 0.005 are marked in cyan; shocks with $\Delta P/P > 0.2$ or $\mathcal{M} \gtrsim 1.08$ are painted yellow (these thresholds, though arbitrary,  are chosen since they appear to cover most of the visible jumps). This figure shows clearly that both mixing and shocks individually provide entropy to the gas (i.e., entropy jumps that are marked exclusively yellow or cyan). Nevertheless, large entropy jumps preferentially occur when both processes work together (i.e., entropy jumps marked with a blended green color). 

Note also that shock heating occurs much more frequently than mixing. Therefore, although heating by each shock is less efficient than each mixing event, as also found by \cite{Hillel16}, it has been unclear whether the {\it cumulative} amount of heating from shocks over the course of the cluster evolution is significant or not. Our simulation is suitable for answering this question because we follow the cluster evolution over billions of years. Furthermore, the strengths and frequencies of shocks are determined by the self-regulated AGN feedback. In Section \ref{sec:contribution} we will present a more quantitative analysis on the relative amounts of heating from different processes.

\subsection{The Eulerian view}
\label{sec:eulerian}

In this section, we investigate the sources of heating from the Eulerian point of view, as if there were static ICM tracer particles. We will first show that the thermodynamic quantities of the gas at any fixed point in space are {\it on average} nearly constant. Together with results derived from the previous section, we argue that advection of heat as well as adiabatic heating/cooling are important within the ambient region/jet cones. 

\begin{figure*}[ptb]
\begin{center}
\includegraphics[scale=0.8]{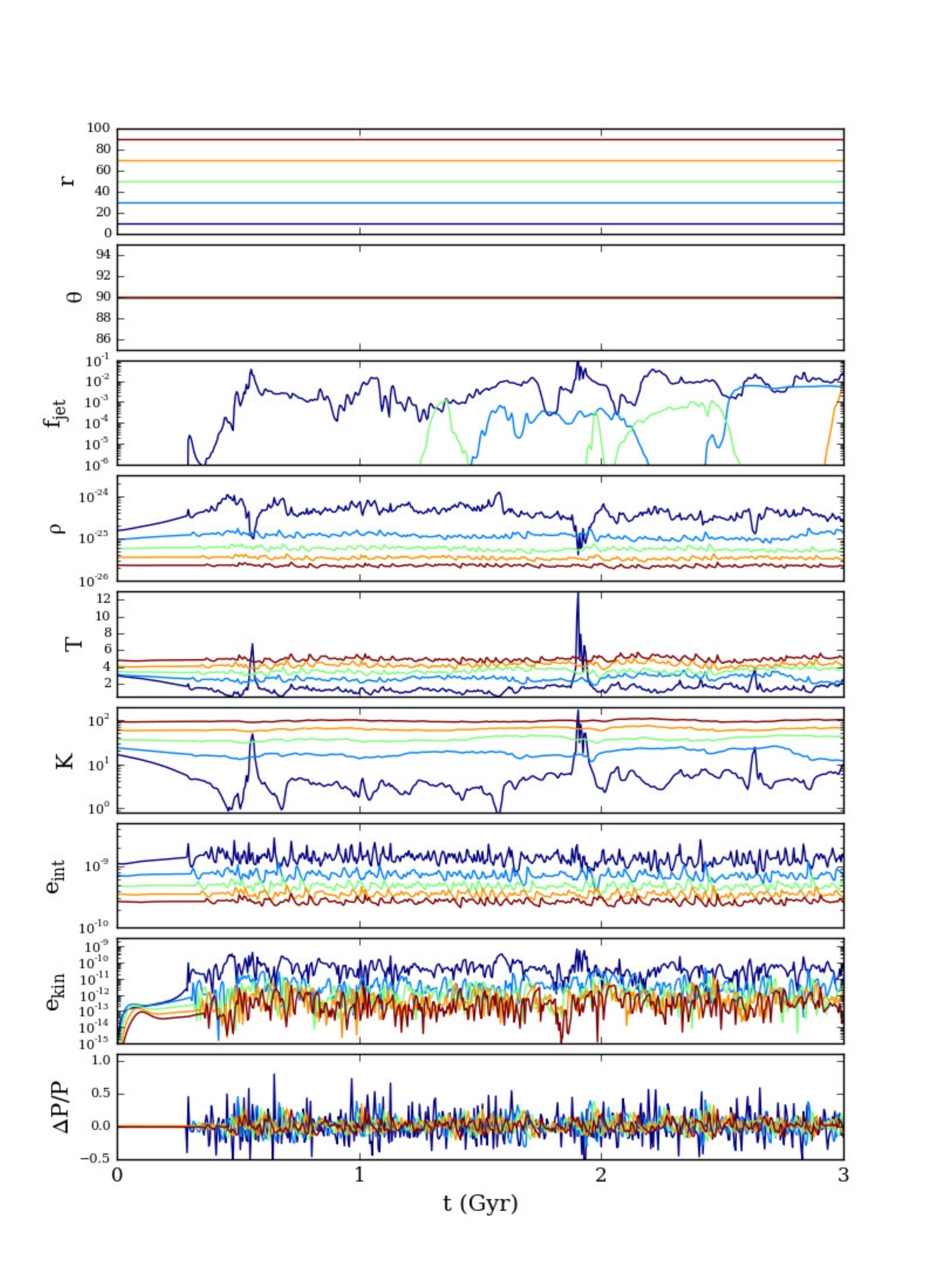} 
\caption{Evolution of hydrodynamic variables at fixed points in space along the $+y$-axis. Meanings of variables are the same as in Figure \ref{fig:TR10}.}
\label{fig:cellx}
\end{center}
\end{figure*}

To study the gas within the ambient region, we plot the relevant hydrodynamic variables at fixed points along the $+y$-axis in Figure \ref{fig:cellx}. In general, all quantities are remarkably steady, with only small fluctuations caused by weak shocks and sound waves. The only exception is the static tracer at $r=10$\ kpc (the navy curve). Since it is very close to the cluster center where cooling time is short and also where jets are launched, greater variations are expected. Nevertheless, aside from the large fluctuations caused by mixing (e.g., $t\sim 0.55, 1.90, 2.60$\ Gyr), the evolution is nearly constant after $t\sim 0.7$\ Gyr (see also the right column of Figure \ref{fig:prf}).   

\begin{figure*}[ptb]
\begin{center}
\includegraphics[scale=0.8]{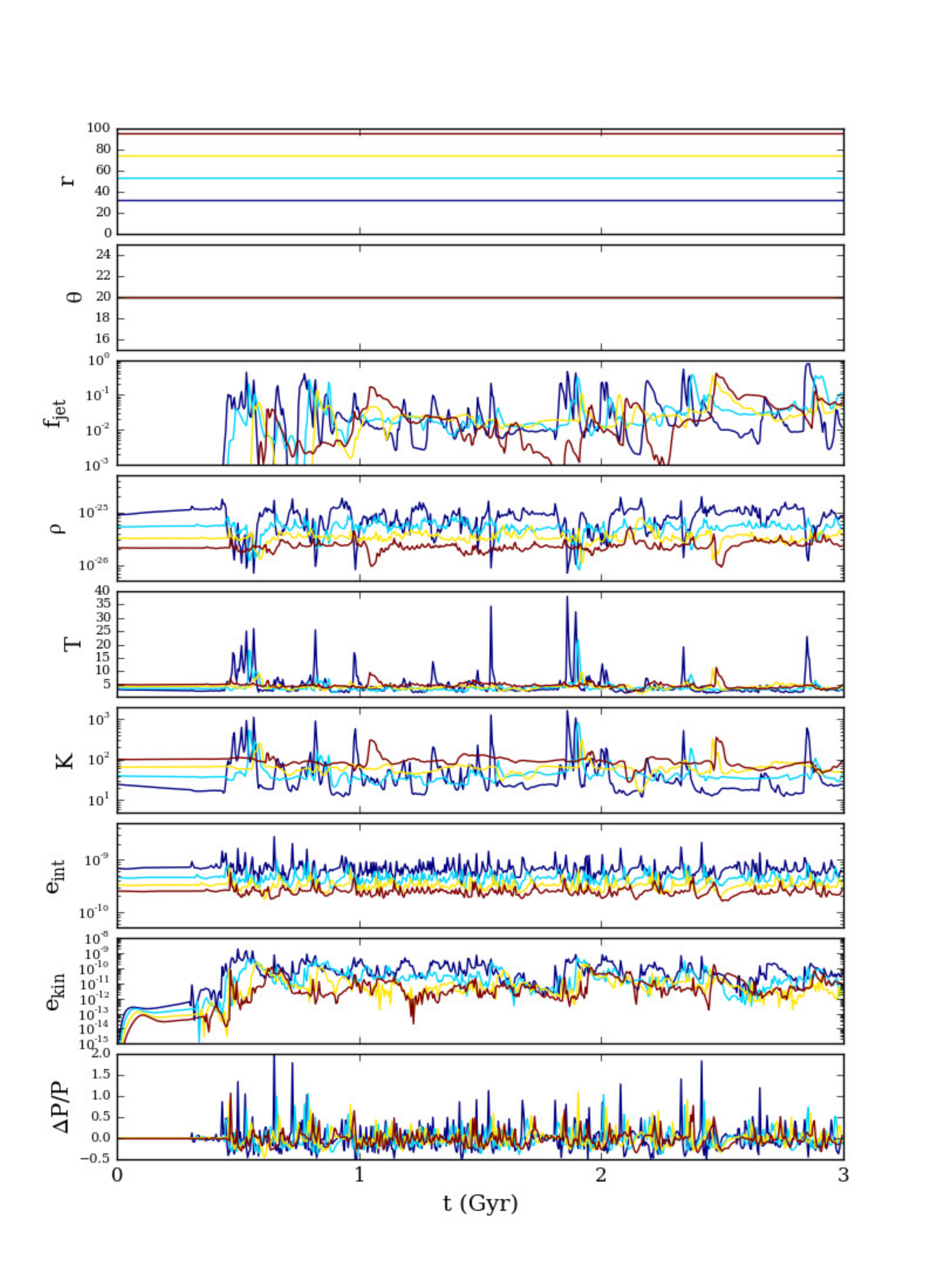} 
\caption{Same as Figure \ref{fig:cellx} but at fixed points along a line of sight $20^\circ$ away from the $+z$-axis. Meanings of variables are the same as in Figure \ref{fig:TR10}.}
\label{fig:cellz20}
\end{center}
\end{figure*}

The quasi-steady nature is also found for gas within the jet cones. Figure \ref{fig:cellz20} shows the hydrodynamic variables at fixed points along a sightline $20^\circ$ away from the $+z$-axis. Since these locations undergo more events of mixing and stronger shocks, they have greater fluctuations than the ambient gas (see Figure \ref{fig:prf}). However, except the excursions caused by mixing and shocks, the gas density, temperature, entropy, and especially internal energy remain essentially unchanged over Gyr timescales. 

How do we reconcile the nearly constant temperature or internal energy with the fact that there is net heating/cooling within the jet cones/ambient region? We can understand this by recalling the internal energy equation:
\begin{equation}
\frac{\partial e_{\rm i}}{\partial t} + \nabla \cdot [(e_{\rm i}+P){\bm v}] - {\bm v}\cdot \nabla{P} = \mathcal{H}({\rm {\bf x}}) - \mathcal{C}({\rm {\bf x}}),
\label{eq:eint}
\end{equation}
where $\mathcal{H}({\rm {\bf x}})$ and $\mathcal{C}({\rm {\bf x}})$ are functions of space. The $\nabla \cdot (e_{\rm i} {\bm v})$ term represents transport of the internal energy by the flow via advection or convection. The part of the equation, $\nabla \cdot (P {\bm v}) - {\bm v} \cdot \nabla P = P(\nabla \cdot {\bm v})$ is the change of internal energy due to adiabatic compression and expansion. From the previous section we have learned that the right-hand-side of the equation is positive/negative within the jet cones/ambient region. In order to make the internal energy roughly constant over time, the net heating/cooling must be balanced by the energy transport and adiabatic energy losses/gains. In the next section, we quantify each term and compare their relative importance.

\subsection{Relative contribution from each process}
\label{sec:contribution}

In the previous sections, we have identified the processes that are relevant for maintaining the quasi-equilibrium state of the cluster, including radiative cooling, bubble mixing, weak shocks, advective and convective transport, and adiabatic compression and expansion. In the following we quantify all these processes and compare their relative importance by computing the volumetric heating and cooling rates. 

In order to construct radial profiles of the heating and cooling rates, we divide the domain within 100 kpc into twenty sectors. First we distinguish the jet cones and the ambient region, and then we further divide them into ten radial bins, i.e., the $i$th bin contains region with $(i-1)\times 10\ {\rm kpc} \le r \le i\times 10\ {\rm kpc}$, where $i \in [1, 10]$. For each process, we compute the heating or cooling `luminosities' (in units of ${\rm erg}\ {\rm s}^{-1}$) in these sectors. The luminosity of a process is the volume integral of the corresponding term in Equation \ref{eq:eint}. More specifically, the luminosities due to cooling, advection/convection, adiabatic compression/expansion, and weak shocks are, respectively,
\begin{eqnarray} 
L_{\rm c} &=& - \int_{V_i} n_{\rm e}^2 \Lambda(T) {\rm d}V, \nonumber \\
L_{\rm [adv, conv]} &=& - \int_{V_i} \nabla \cdot {\bm Q}_{\rm[adv, conv]} {\rm d}V = - \oint_{S_i} Q_{\rm [adv,conv]} {\rm d}S, \nonumber \\
L_{\rm ad} &=& - \oint_{S_i} P{\bm v} \cdot {\rm d}{\bm S} + \int_{V_i} {\bm v} \cdot \nabla P {\rm d}V, \nonumber \\
L_{\rm sh} &=& \int_{V_i} \frac{{\rm d}H}{{\rm d}t} {\rm d}V,
\label{eq:emis}
\end{eqnarray}
where $V_i$ and $S_i$ are the volume and surface of each sector and ${\rm d}H/{\rm d}t$ is the heating rate by weak shocks (see end of Section \ref{sec:method}). The advective and convective fluxes are defined as
\begin{eqnarray}
Q_{\rm adv} &=& \frac{1}{\gamma-1} k_{\rm B} ( \langle n\rangle \langle T \rangle \langle v_\perp \rangle + \langle T \rangle \langle \delta n \delta v_\perp \rangle), \nonumber \\
Q_{\rm conv} &=& \frac{1}{\gamma-1} k_{\rm B} (\langle n \rangle \langle \delta v_\perp \delta T \rangle + \langle v_\perp \rangle \langle \delta n \delta T \rangle \nonumber \\
&+& \langle \delta n \delta T \delta v_\perp \rangle),
\label{eq:flux}
\end{eqnarray}
where $v_\perp$ is the velocity component perpendicular the surface of integration, angle brackets denote averages over the surfaces, and $\delta$ refers to the local deviation from the mean value of a quantity, e.g., $\delta n \equiv n - \langle n \rangle$. The volumetric heating or cooling rates, or `emissivities' (in units of ${\rm erg}\ {\rm s}^{-1}\ {\rm cm}^{-3}$), are then calculated by dividing the luminosities by the volume of each sector, $V_i$. The luminosity and emissivity profiles for both the jet cones and the ambient region are obtained in this fashion for every simulation output. Note that all the relevant processes could be directly estimated in the way described above, except for heating from mixing. Since the process of mixing is a complex fluid dynamic problem and no simple models exist for our current application, we will infer the amount of mixing heating from the fact that there is rough balance between heating and cooling within the jet cones. 

\begin{figure*}[ptb]
\begin{center}
\includegraphics[scale=0.7]{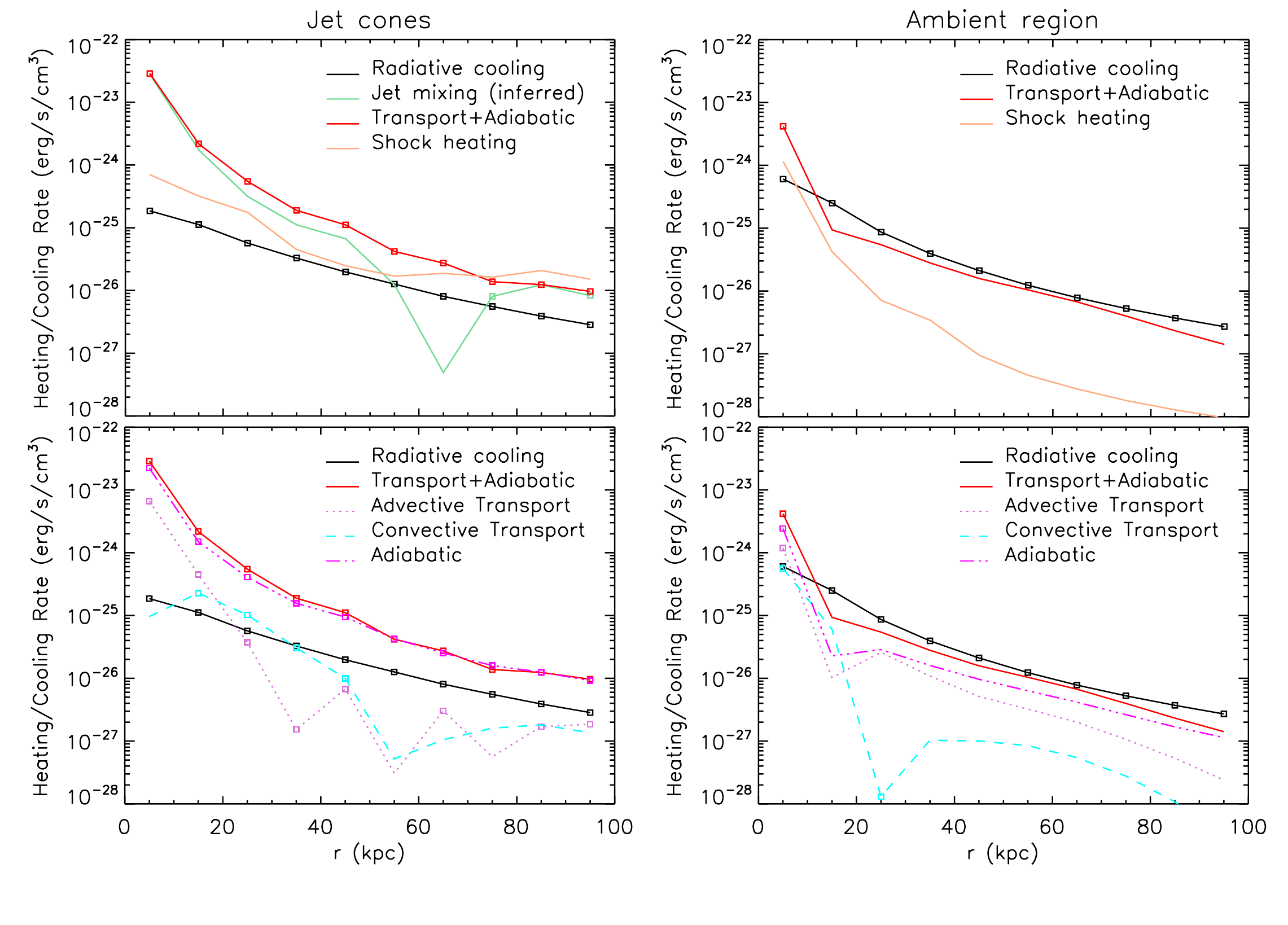} 
\caption{Time-averaged (between 1 and 3 Gyr) radial profiles of heating and cooling emissivities in units of ${\rm erg}\ {\rm s}^{-1}\ {\rm cm}^{-3}$ for the jet cones (left column) and the ambient region (right column). Data marked with open squares represent processes that are cooling for a certain radial bin. The top panels shows the contributions from radiative cooling, shock heating, and transport plus adiabatic processes. The last term is further broken down into advection, convection, and adiabatic compression/expansion in the bottom panels. The amount of mixing heating shown in the top left panel is inferred from the fact that there is rough balance between heating and cooling (see Figure \ref{fig:cellz20}).}
\label{fig:emis}
\end{center}
\end{figure*}

Figure \ref{fig:emis} shows the time-averaged profiles of heating and cooling emissivities for the jet cones (left column) and the ambient region (right column). The profiles are averaged between $t=1$ and 3 Gyr in order to study the heating and cooling balance when the cluster is in the quasi-equilibrium state. Within the jet cones, we find that because the gas is relatively underdense, radiative cooling is not the dominant cooling process. Instead, cooling due to transport and adiabatic processes is more significant. By breaking down the contributions further (see the bottom left panel), we find that the energy losses are primarily due to adiabatic expansion, while energy transport by advection and convection is sub-dominant. In terms of heating, the amount of heat provided by repetitive weak shocks is enough to offset the radiative losses, but not the transport plus adiabatic losses (except for larger radii, i.e., $r\gtrsim 80$\ kpc). In order to maintain the ICM in quasi-equilibrium (see Figure \ref{fig:cellz20}), heating from bubble mixing is needed to balance the substantial energy losses from transport and adiabatic processes. We derived such an inferred profile for bubble mixing and plotted it in the upper left panel. As can be seen from the figure, bubble mixing appears to be a more important heating source than weak shocks within the jet cones, especially within the inner 80 kpc. This is consistent with the expectation that mixing is more efficient than shocks (see Section \ref{sec:lagrangian}). However, here we quantitatively show that the {\it cumulative} amount of heat provided by mixing is also greater than weak shocks despite of their high frequency. Note that the emissivities shown here does not exclude the bubbles. However, we repeated the analyses by computing all quantities in Equations \ref{eq:emis} and \ref{eq:flux} weighted by $f_{\rm gas}$ and the conclusions are unchanged. This implies that adiabatic expansion is associated with not only the bubbles, but also the uplifted ICM as they move outward. This is consistent with the density evolution of the ICM tracer particles as seen in Figures \ref{fig:TR10} and \ref{fig:TR30}.   

The emissivity components for the ambient region are plotted in the right columns of Figure \ref{fig:emis}.  As have been alluded to in the previous sections, there is net cooling within the ambient region, i.e., the only available heating source is weak shocks, and it is insufficient to overcome radiative cooling. What maintains the ambient gas at a given radius to be quasi-steady (see Figure \ref{fig:cellx}) is energy from transport and adiabatic processes. As shown in the bottom right panel, adiabatic compression and advection are both non-negligible (the former has a slightly more contribution), while convection plays a lesser role. By estimating the amount of advective energy transfer through the surfaces of each sector, we find that the energy primarily flows from the interface between the jet cones and the ambient region, as opposed to radial inflow from larger radii. In other words, the ambient region, though having net cooling, is pumped by flows of energy from already-heated gas. The gas further gains energy by adiabatic compression as it flows inward radially as a part of gentle circulation. Note that for the first two radial bins in the upper right panel of Figure \ref{fig:emis}, there appears to be some net energy losses. This loss of internal energy could be compensated by mixing with the non-negligible amount of jet materials within the inner 20 kpc in the ambient region (see the right columns of Figure \ref{fig:slices}). 

In terms of the overall contributions by different processes within the whole cluster core (i.e., the averaged heating and cooling luminosities integrated within 100 kpc), we find that within the jet cones, $L_{\rm c}:L_{\rm mix}:(L_{\rm adv}+L_{\rm conv}+L_{\rm ad}):L_{\rm sh}=(-1.52\times10^{44}):9.14\times 10^{44}:(-1.37\times 10^{45}):4.08\times 10^{44} \simeq (-0.11): 0.67:( -1.0): 0.30$. That is, cooling is dominated by adiabatic expansion instead of radiative cooling; heating from bubble mixing contributes about twice the amount of heating from shocks within the cooling radius. For the ambient region, $L_{\rm c}:(L_{\rm adv}+L_{\rm conv}+L_{\rm ad}):L_{\rm sh}=(-1.20\times10^{45}):9.95\times 10^{44}:2.05\times 10^{44} \simeq (-1.0): 0.83: 0.17$. In other words, shock heating slows down radiative cooling by $\sim 17\%$, while the rest of gas internal energy comes from advection and adiabatic compression.  The overall energetics suggest that in our simulation, most injected energy from the AGN is stored within the bubbles, whereas the energy associated with shocks contain a minor portion. This is consistent with the energy partitions inferred by recent analysis of the Perseus cluster \citep{Zhuravleva16}.

   

\section{Discussion}
\label{sec:discussion}

\subsection{Turbulent heating}
\label{sec:turb}

\begin{figure}[ptb]
\begin{center}
\includegraphics[scale=0.6]{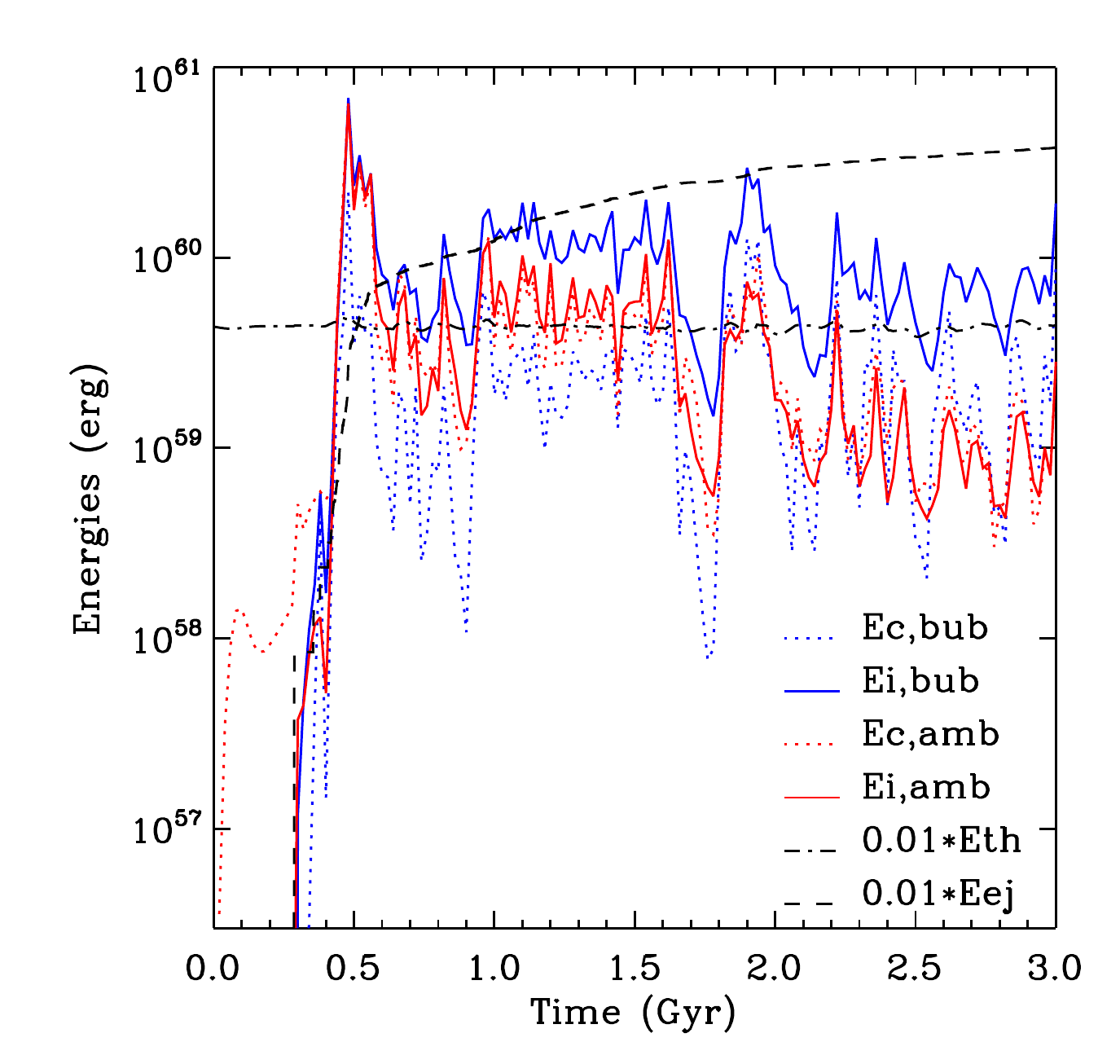} 
\caption{Evolution of the kinetic energies of the compressible (dotted) and incompressible (solid) velocity fields, which are further separated by regions containing the bubbles ($f_{\rm jet} \ge 0.01$; blue) and excluding the bubbles ($f_{\rm jet} < 0.01$; red). Dashed and dashed-dotted lines show one percent of the injected energy from the AGN and the total gas thermal energy within 100\ kpc, respectively.}
\label{fig:ekin}
\end{center}
\end{figure}

In Section \ref{sec:lagrangian}, we showed that the kinetic energy associated with the tracer particles is always only at the percent level compared to the internal energy (Figure \ref{fig:TR10} and \ref{fig:TR30}). The kinetic energy contains contributions not only from turbulence but also from shocks and waves. To better understand the energetics, we decompose the velocity fields into the compressible (which traces shocks and waves) and incompressible (which measures turbulence and $g$-modes) components \citep{Yang15, Reynolds15}. The total kinetic energies within 100 kpc for regions containing the bubbles ($f_{\rm jet} \ge 0.01$) and excluding the bubbles ($f_{\rm jet} < 0.01$) are plotted in Figure \ref{fig:ekin}. For comparison, the injected AGN energy multiplied by 0.01 is overplotted.   

We find that the turbulent energy contained within the bubbles is the largest among all components, $\sim 2-3\%$ of the gas thermal energy. However, all kinetic energies are only about one percent of the gas thermal energy or the injected feedback energy of the AGN, suggesting that the driving of turbulence by AGN feedback is {\it inefficient}. This result is broadly consistent with the recent simulations with more idealized AGN energy injections by \cite{Reynolds15} and \cite{Hillel16}. As discussed in \cite{Yang15}, the inefficiency of AGN-driven turbulence is due to the facts that the typical frequency of AGN outbursts is higher than that required to excite $g$-waves, and that the $g$-modes are likely trapped within a small volume within the CC.   

Given the small amount of energy available for turbulent cascade, we conclude that turbulent heating is not the primary source of heating in our simulated cluster. This is in tension with recent analyses of the observational data for the Perseus cluster \citep{Zhuravleva14, Zhuravleva16}, which suggests that turbulent energy is $\sim 10\%$ of the gas thermal energy and the estimated timescale for turbulence dissipation is comparable to what is required to compensate radiative cooling losses. More efforts, both theoretically and observationally, are needed to resolve this issue. It remains a possibility that predicting the level of AGN-driven turbulence requires models beyond pure hydrodynamics \citep{Reynolds15}. 
Observationally, a glimpse at the turbulent dynamics of the ICM will be given by the ASTRO-H/Hitomi observation of the core of the Perseus cluster obtained prior to the loss of that mission in March 2016 \citep{Takahashi16}. A more complete observational characterization of ICM turbulence must wait for a future X-ray micro-calorimeter mission.
If future works prove both statements true (i.e., observed turbulence within cluster cores is energetically significant but AGN driving is inefficient), then it would imply that turbulence within the cores originates primarily from other mechanisms such as sloshing \citep[e.g.,][]{Zuhone13}, galaxy motions \citep[e.g.,][]{Ruszkowski11a}, and cascades from large-scale turbulence generated by mergers \citep[e.g.,][]{Vazza13}. Note that these are not `feedback' mechanisms as they are associated with the dynamical state of the whole cluster, and therefore it may still be challenging to solve the cooling-flow problem via turbulent heating. Alternatively, it might be that AGN feedback is still the solution to cooling catastrophes (e.g., by mixing and shocks), but the dissipation of turbulence is somehow significantly suppressed due to poorly understood plasma physics on microscopic scales. For example, the fact that dissipation of magnetohydrodynamic turbulence is highly localized both spatially and temporally requires heat to diffuse faster than being radiated away \citep{Zhdankin15}. In the regime where the ICM is weakly collisional, the Braginskii viscosity only dissipates certain components of the velocity gradients \citep{Schekochihin07}, and thus the available energy for turbulence dissipation depends on the partitions between different (compressible versus incompressible) modes \citep{Miniati15}. To address these questions is beyond the scope of this paper and will be parts of our future work.    

\subsection{Limitations of current simulation}
\label{sec:limitation}

In this work, we intentionally focused on the purely hydrodynamic effects of AGN heating, while magnetic fields and plasma transport processes of the ICM are omitted. The details of mixing between the bubbles and the ambient ICM may be affected by the magnetic fields \citep[e.g.,][]{Robinson04, Ruszkowski08} and viscosity \citep{Reynolds05}. Also, some of the potential heating mechanisms are not explored in the current study. For example, while sound waves in our simulation simply propagate out of the core, their energy can be dissipated within a conducting and/or viscous medium \citep[e.g.,][]{Bruggen05, Fabian05}. Cavity heating \citep[e.g.,][]{Churazov01} is also not included in our analysis, since the gravitational energy gained by the uplifted gas is transformed into kinetic energy when it refills the wakes of the bubbles. In principle this kinetic energy could be viscously dissipated into heat, but we do not explicitly include viscosity in the current simulations. These additional mechanisms will change the relative partitions among different heating mechanisms. We will incorporate these processes in our future simulations in order to quantify their importance. Note, however, that both cavity heating and sound-wave heating come from the kinetic energy of the gas (within the wakes and sound waves, respectively), which appears to be subdominant in terms of the total feedback energy (see Figure \ref{fig:ekin}). Therefore, the current study suggests that these two mechanisms may not be energetically dominant. 

The current simulation have neglected externally-driven turbulence in order to probe perturbations solely due to the AGN. In reality, turbulence generated by other sources \citep[e.g.,][]{Ruszkowski11a, Zuhone13, Vazza13} could co-exist with the AGN. The pre-existing turbulent motions are expected to displace the buoyantly rising bubbles \citep[e.g.,][]{Morsony10} so that the nearly-isotropic distributions of the observed bubbles can be better reproduced \citep{Fabian06}. In this case, the heating due to mixing between the bubbles and the ICM would not be confined within the jet cones, but would be distributed more isotropically. Therefore, instead of large-scale circulation consisting of a meridian outflow and an equatorial inflow \citep{Soker15, Hillel16}, the ICM within the CCs may resemble a pot of boiling waters, with outwardly rising gas parcels directly heated by the AGN and under-heated gas inter-penetrating between bubbles and sinking toward the center \citep{Lim08}. 
In any case, our current simulation has shown the success of AGN feedback even when the volume of heating is limited. Including external turbulence should only aid the AGN by the additional transport of heat.    

Our simulation, as well as recent simulations of AGN feedback that aim to reproduce the properties of the CCs and TI \citep[e.g.,][]{Gaspari11, Li14, Prasad15}, have been focusing on momentum-driven feedback. Since the kinetic energy of the jets is quickly thermalized by shocks, one implicit assumption of these simulations is that the bubbles are filled with ultra-hot (but non-relativistic) gas. Therefore, heating the ICM by mixing is relatively easy since the bubbles contain mainly thermal energy. In reality, the composition of the radio bubbles is still uncertain. Future high-resolution observations of the Sunyaev-Zel'dovich effect of cavities could potentially constrain the bubble composition \citep{Pfrommer05}. If relativistic particles dominate the bubble pressure, i.e., if the bubbles are inflated by cosmic-ray dominated jets \citep{Guo11}, heating by mixing would likely be less efficient \citep{Mathews08}. The details of this scenario, e.g., the mechanisms of heating, the interaction between cosmic-ray filled bubbles and the surrounding ICM, and how these processes depend on plasma microphysics, require further investigation. We will address some of these issues in our forthcoming papers. 


\section{Conclusions}
\label{sec:conclusion}

The key to solving the cooling-flow problem of galaxy clusters is most likely linked to AGN feedback. However, how exactly the feedback energy is transformed into heat, in particular, isotropically, remains unclear. In this work, we use a representative hydrodynamic simulation of a CC cluster with self-regulated, momentum-driven AGN feedback to gain insights into the relative contributions from various physical mechanisms, including radiative cooling, shock heating, bubble mixing, fluid motions (advection, convection, and turbulence), and adiabatic processes.  

The picture of AGN feedback drawn from our simulation is summarized as follows. The AGN jets inflate bubbles and produce weak shocks, which increase entropy of the ICM by shocks and mixing with the hot bubble materials. The heat is primarily dumped within the jet cones ($\lesssim 30^\circ$ away from the $z$-axis); however, the Eulerian internal energy of the gas is not dramatically increased because the gas adiabatically expands as it is uplifted by the buoyantly rising bubbles. A gentle circulation over multiple billions of years is established by the episodic AGN activity, where there is net gas outflow within the jet cones, and some of the gas can refill the ambient region and eventually move inward toward the cluster center. Within the ambient region, heating from weak shocks slows down the cooling losses but not completely, and hence forming a `reduced' cooling flow. However, the thermodynamic properties of the ambient ICM stay nearly constant over Gyr timescales because the ambient region is replenished by previously-heated gas from the jet cones, and also because the gas is adiabatically compressed as it moves inward.   


In terms of X-ray emission, the dominating component is from the reduced cooling flow, whereas the heated gas mixed with the bubbles is relatively underdense and less luminous. This allows the kinetic AGN feedback model to explain the positive temperature gradients of observed CCs without producing signatures of over-heated gas in the vicinity of bubbles. This aspect of our model is qualitatively similar to the one-dimensional circulation model proposed by \cite{Mathews04}, though the mechanisms of heating were not specified in their work. We also confirmed that AGN heating occurs inside-out (Figure \ref{fig:emis}), which is an essential ingredient to explain the slope of the soft X-ray spectrum in the cold accretion plus momentum-driven feedback scenario \citep{Gaspari15b}.

Among the sources of heat that we explored, we find that both bubble mixing and shock heating are important, especially within the jet cones. Moreover, the amount of heat provided by mixing is greater than by shocks. There has been a long debate in the community about whether bubble mixing or shock heating is the dominant heat source. On one hand, observations that measure the shock strengths and frequencies suggest enough heating from weak shocks to counteract radiative cooling \citep[e.g.,][]{Randall15}. On the other hand, simulations using tracer fluids found more efficient heating by bubble mixing than shocks \citep[e.g.,][]{Hillel16}. Interestingly, we find that the two seemingly contradictory statements are both true. Indeed, the cumulative amount of heating from repetitive shocks turns out to be comparable to the radiative losses. However, greater energy losses are associated with adiabatic expansion of the gas, which is {\it invisible} from the observed X-ray images. This adiabatic cooling losses have to be compensated by mixing heating. 

For the simulated cluster, the kinetic energy of the incompressible velocity field (which mainly traces turbulence) is only at the percent level compared to the injected feedback energy (Figure \ref{fig:ekin}), suggesting that turbulent heating is not the dominant source of heating in our simulation. The implications of this result are discussed in Section \ref{sec:turb}.  

One of the biggest puzzles of AGN feedback in CCs is, given that AGN jets are directional, how to heat the whole cluster core isotropically. Specifically, previous works have been looking for sources of heating that can balance {\it radiative cooling} at nearly {\it every location} within the cluster core. Such requirement poses great challenges to all the proposed heating mechanisms since none of them can easily provide heating with radial and azimuthal profiles similar to that of radiative cooling. This difficulty can be overcome by taking into account energy transport and adiabatic processes, which are natural consequences of buoyantly rising bubbles. The results of our simulation suggest that the AGN jets do not need to deposit energy isotropically. Instead, they provide excess entropy to parts of the core (i.e., the jet cones in our setup) by bubble mixing and weak shocks. The excess entropy is spent both on heating and adiabatic expansion of the gas. Through the process of gentle circulation, the heated gas brings energy into the under-heated parts of the core. This advection, together with adiabatic compression, provide energy that balances the net entropy losses for the under-heated gas. In other words, the fluid dynamics self-adjusts such that it compensates and transports the heat provided by the AGN, and thus no fine-tuning of the heating profiles is needed. 

Our model suggests that self-regulated, momentum-driven AGN feedback, averaged over time, does not provide heat that balances radiative cooling at every location within the cluster core, i.e., the ICM is {\it not} in a global thermal equilibrium. Instead, there can be a significant variation in $\mathcal{H}-\mathcal{C}$ (of magnitude comparable to $\mathcal{C}$) within the core. This result has important implications for studies of TI within cluster cores, which often assume global thermal equilibrium with small local perturbations \citep{McCourt12, Sharma12, Meece15}. In particular, the properties of TI are expected to be different for a background medium that is in global thermal equilibrium versus a (reduced) cooling flow \citep{Sharma10}. Therefore, the former assumption may not be adequate, but the dynamic nature of the ICM must be taken into account in order to properly model TI and multiphase gas within the CCs \citep[e.g.,][]{Gaspari12, Li14}.     

The current simulation has several limitations, including the assumptions of pure hydrodynamics, the absence of external turbulence, and thermal-pressure dominated bubbles (see Section \ref{sec:limitation} for detailed discussions). Future simulations including processes of increased levels of complexity, such as magnetic fields, anisotropic thermal conduction, Braginskii viscosity, external turbulence, and cosmic rays, will further improve our understanding of the intricate balance between heating and cooling within cluster cores. 


\acknowledgments

The authors would like to thank Yuan Li, Noam Soker, Massimo Gaspari, and an anonymous referee for helpful discussion and comments that helped to improve the manuscript. HYKY acknowledges support by NASA through Einstein Postdoctoral Fellowship grant number PF4-150129 awarded by the Chandra X-ray Center, which is operated by the Smithsonian Astrophysical Observatory for NASA under contract NAS8-03060. CSR is grateful for financial support from the Simons Foundation (through a Simons Fellowship in Theoretical Physics) and the National Science Foundation (under grant AST1333514).  The simulations presented in this paper were performed on the {\tt Deepthought2} cluster, maintained and supported by the Division of Information Technology at the University of Maryland College Park. FLASH was developed largely by the DOE-supported ASC/Alliances Center for Astrophysical Thermonuclear Flashes at the University of Chicago.  


\bibliography{agn}

\begin{thebibliography}{96}
\expandafter\ifx\csname natexlab\endcsname\relax\def\natexlab#1{#1}\fi

\bibitem[{{Allen} {et~al.}(2001{\natexlab{a}}){Allen}, {Fabian}, {Johnstone},
  {Arnaud}, \& {Nulsen}}]{Allen01}
{Allen}, S.~W., {Fabian}, A.~C., {Johnstone}, R.~M., {Arnaud}, K.~A., \&
  {Nulsen}, P.~E.~J. 2001{\natexlab{a}}, \mnras, 322, 589

\bibitem[{{Allen} {et~al.}(2001{\natexlab{b}}){Allen}, {Schmidt}, \&
  {Fabian}}]{Allen01b}
{Allen}, S.~W., {Schmidt}, R.~W., \& {Fabian}, A.~C. 2001{\natexlab{b}},
  \mnras, 328, L37

\bibitem[{{Best} {et~al.}(2007){Best}, {von der Linden}, {Kauffmann},
  {Heckman}, \& {Kaiser}}]{Best07}
{Best}, P.~N., {von der Linden}, A., {Kauffmann}, G., {Heckman}, T.~M., \&
  {Kaiser}, C.~R. 2007, \mnras, 379, 894

\bibitem[{{B{\^\i}rzan} {et~al.}(2012){B{\^\i}rzan}, {Rafferty}, {Nulsen},
  {et~al.}}]{Birzan12}
{B{\^\i}rzan}, L., {Rafferty}, D.~A., {Nulsen}, P.~E.~J., {et~al.} 2012,
  \mnras, 427, 3468

\bibitem[{{Blanton} {et~al.}(2009){Blanton}, {Randall}, {Douglass},
  {et~al.}}]{Blanton09}
{Blanton}, E.~L., {Randall}, S.~W., {Douglass}, E.~M., {et~al.} 2009, \apjl,
  697, L95

\bibitem[{{Bregman} \& {David}(1988)}]{Bregman88}
{Bregman}, J.~N., \& {David}, L.~P. 1988, \apj, 326, 639

\bibitem[{{Brighenti} \& {Mathews}(2002)}]{Brighenti02}
{Brighenti}, F., \& {Mathews}, W.~G. 2002, \apj, 573, 542

\bibitem[{{Br{\"u}ggen}(2003)}]{Bruggen03}
{Br{\"u}ggen}, M. 2003, \apj, 592, 839

\bibitem[{{Br{\"u}ggen} {et~al.}(2005){Br{\"u}ggen}, {Ruszkowski}, \&
  {Hallman}}]{Bruggen05}
{Br{\"u}ggen}, M., {Ruszkowski}, M., \& {Hallman}, E. 2005, \apj, 630, 740

\bibitem[{{Bryan} {et~al.}(2014){Bryan}, {Norman}, {O'Shea},
  {et~al.}}]{Bryan14}
{Bryan}, G.~L., {Norman}, M.~L., {O'Shea}, B.~W., {et~al.} 2014, \apjs, 211, 19

\bibitem[{{Burns} {et~al.}(2008){Burns}, {Hallman}, {Gantner}, {Motl}, \&
  {Norman}}]{Burns08}
{Burns}, J.~O., {Hallman}, E.~J., {Gantner}, B., {Motl}, P.~M., \& {Norman},
  M.~L. 2008, \apj, 675, 1125

\bibitem[{{Cattaneo} \& {Teyssier}(2007)}]{Cattaneo07}
{Cattaneo}, A., \& {Teyssier}, R. 2007, \mnras, 376, 1547

\bibitem[{{Cavagnolo} {et~al.}(2008){Cavagnolo}, {Donahue}, {Voit}, \&
  {Sun}}]{Cavagnolo08}
{Cavagnolo}, K.~W., {Donahue}, M., {Voit}, G.~M., \& {Sun}, M. 2008, \apjl,
  683, L107

\bibitem[{{Cavagnolo} {et~al.}(2009){Cavagnolo}, {Donahue}, {Voit}, \&
  {Sun}}]{Cavagnolo09}
---. 2009, \apjs, 182, 12

\bibitem[{{Churazov} {et~al.}(2001){Churazov}, {Br{\"u}ggen}, {Kaiser},
  {B{\"o}hringer}, \& {Forman}}]{Churazov01}
{Churazov}, E., {Br{\"u}ggen}, M., {Kaiser}, C.~R., {B{\"o}hringer}, H., \&
  {Forman}, W. 2001, \apj, 554, 261

\bibitem[{{David} {et~al.}(2001){David}, {Nulsen}, {McNamara},
  {et~al.}}]{David01}
{David}, L.~P., {Nulsen}, P.~E.~J., {McNamara}, B.~R., {et~al.} 2001, \apj,
  557, 546

\bibitem[{{Dennis} \& {Chandran}(2005)}]{Dennis05}
{Dennis}, T.~J., \& {Chandran}, B.~D.~G. 2005, \apj, 622, 205

\bibitem[{{Dubey} {et~al.}(2008){Dubey}, {Reid}, \& {Fisher}}]{Dubey08}
{Dubey}, A., {Reid}, L.~B., \& {Fisher}, R. 2008, Physica Scripta, T132, 014046

\bibitem[{{Dubois} {et~al.}(2010){Dubois}, {Devriendt}, {Slyz}, \&
  {Teyssier}}]{Dubois10}
{Dubois}, Y., {Devriendt}, J., {Slyz}, A., \& {Teyssier}, R. 2010, \mnras, 409,
  985

\bibitem[{{Dunn} \& {Fabian}(2008)}]{Dunn08}
{Dunn}, R.~J.~H., \& {Fabian}, A.~C. 2008, \mnras, 385, 757

\bibitem[{{Edge}(2001)}]{Edge01}
{Edge}, A.~C. 2001, \mnras, 328, 762

\bibitem[{{Fabian}(1994)}]{Fabian94}
{Fabian}, A.~C. 1994, ARA\&A, 32, 277

\bibitem[{{Fabian} {et~al.}(2005){Fabian}, {Reynolds}, {Taylor}, \&
  {Dunn}}]{Fabian05}
{Fabian}, A.~C., {Reynolds}, C.~S., {Taylor}, G.~B., \& {Dunn}, R.~J.~H. 2005,
  \mnras, 363, 891

\bibitem[{{Fabian} {et~al.}(2003){Fabian}, {Sanders}, {Allen},
  {et~al.}}]{Fabian03}
{Fabian}, A.~C., {Sanders}, J.~S., {Allen}, S.~W., {et~al.} 2003, \mnras, 344,
  L43

\bibitem[{{Fabian} {et~al.}(2000){Fabian}, {Sanders}, {Ettori},
  {et~al.}}]{Fabian00}
{Fabian}, A.~C., {Sanders}, J.~S., {Ettori}, S., {et~al.} 2000, \mnras, 318,
  L65

\bibitem[{{Fabian} {et~al.}(2006){Fabian}, {Sanders}, {Taylor},
  {et~al.}}]{Fabian06}
{Fabian}, A.~C., {Sanders}, J.~S., {Taylor}, G.~B., {et~al.} 2006, \mnras, 366,
  417

\bibitem[{{Fryxell} {et~al.}(2000){Fryxell}, {Olson}, {Ricker},
  {et~al.}}]{Flash}
{Fryxell}, B., {Olson}, K., {Ricker}, P., {et~al.} 2000, \apjs, 131, 273

\bibitem[{{Gaspari}(2015)}]{Gaspari15b}
{Gaspari}, M. 2015, \mnras, 451, L60

\bibitem[{{Gaspari} {et~al.}(2015){Gaspari}, {Brighenti}, \&
  {Temi}}]{Gaspari15}
{Gaspari}, M., {Brighenti}, F., \& {Temi}, P. 2015, \aap, 579, A62

\bibitem[{{Gaspari} {et~al.}(2011){Gaspari}, {Melioli}, {Brighenti}, \&
  {D'Ercole}}]{Gaspari11}
{Gaspari}, M., {Melioli}, C., {Brighenti}, F., \& {D'Ercole}, A. 2011, \mnras,
  411, 349

\bibitem[{{Gaspari} {et~al.}(2012){Gaspari}, {Ruszkowski}, \&
  {Sharma}}]{Gaspari12}
{Gaspari}, M., {Ruszkowski}, M., \& {Sharma}, P. 2012, \apj, 746, 94

\bibitem[{{Guo} \& {Mathews}(2011)}]{Guo11}
{Guo}, F., \& {Mathews}, W.~G. 2011, \apj, 728, 121

\bibitem[{{Guo} \& {Oh}(2008)}]{Guo08}
{Guo}, F., \& {Oh}, S.~P. 2008, \mnras, 384, 251

\bibitem[{{Hillel} \& {Soker}(2016)}]{Hillel16}
{Hillel}, S., \& {Soker}, N. 2016, \mnras, 455, 2139

\bibitem[{{Hoffer} {et~al.}(2012){Hoffer}, {Donahue}, {Hicks}, \&
  {Barthelemy}}]{Hoffer12}
{Hoffer}, A.~S., {Donahue}, M., {Hicks}, A., \& {Barthelemy}, R.~S. 2012,
  \apjs, 199, 23

\bibitem[{{Hudson} {et~al.}(2010){Hudson}, {Mittal}, {Reiprich},
  {et~al.}}]{Hudson10}
{Hudson}, D.~S., {Mittal}, R., {Reiprich}, T.~H., {et~al.} 2010, \aap, 513, A37

\bibitem[{{Kim} \& {Narayan}(2003)}]{Kim03}
{Kim}, W.-T., \& {Narayan}, R. 2003, \apjl, 596, L139

\bibitem[{{Lakhchaura} {et~al.}(2016){Lakhchaura}, {Deep Saini}, \&
  {Sharma}}]{Lakhchaura16}
{Lakhchaura}, K., {Deep Saini}, T., \& {Sharma}, P. 2016, arXiv: 1601.02347

\bibitem[{{Li} \& {Bryan}(2012)}]{Li12}
{Li}, Y., \& {Bryan}, G.~L. 2012, \apj, 747, 26

\bibitem[{{Li} \& {Bryan}(2014)}]{Li14}
---. 2014, \apj, 789, 153

\bibitem[{{Li} {et~al.}(2015){Li}, {Bryan}, {Ruszkowski}, {Voit}, {O'Shea}, \&
  {Donahue}}]{Li15}
{Li}, Y., {Bryan}, G.~L., {Ruszkowski}, M., {Voit}, G.~M., {O'Shea}, B.~W., \&
  {Donahue}, M. 2015, \apj, 811, 73

\bibitem[{{Lim} {et~al.}(2008){Lim}, {Ao}, \& {Dinh-V-Trung}}]{Lim08}
{Lim}, J., {Ao}, Y., \& {Dinh-V-Trung}. 2008, \apj, 672, 252

\bibitem[{{Mathews} \& {Brighenti}(2008)}]{Mathews08}
{Mathews}, W.~G., \& {Brighenti}, F. 2008, \apj, 685, 128

\bibitem[{{Mathews} {et~al.}(2004){Mathews}, {Brighenti}, \&
  {Buote}}]{Mathews04}
{Mathews}, W.~G., {Brighenti}, F., \& {Buote}, D.~A. 2004, \apj, 615, 662

\bibitem[{{Mathews} {et~al.}(2006){Mathews}, {Faltenbacher}, \&
  {Brighenti}}]{Mathews06}
{Mathews}, W.~G., {Faltenbacher}, A., \& {Brighenti}, F. 2006, \apj, 638, 659

\bibitem[{{McCourt} {et~al.}(2012){McCourt}, {Sharma}, {Quataert}, \&
  {Parrish}}]{McCourt12}
{McCourt}, M., {Sharma}, P., {Quataert}, E., \& {Parrish}, I.~J. 2012, \mnras,
  419, 3319

\bibitem[{{McDonald} {et~al.}(2011{\natexlab{a}}){McDonald}, {Veilleux}, \&
  {Mushotzky}}]{McDonald11a}
{McDonald}, M., {Veilleux}, S., \& {Mushotzky}, R. 2011{\natexlab{a}}, \apj,
  731, 33

\bibitem[{{McDonald} {et~al.}(2011{\natexlab{b}}){McDonald}, {Veilleux},
  {Rupke}, {Mushotzky}, \& {Reynolds}}]{McDonald11b}
{McDonald}, M., {Veilleux}, S., {Rupke}, D.~S.~N., {Mushotzky}, R., \&
  {Reynolds}, C. 2011{\natexlab{b}}, \apj, 734, 95

\bibitem[{{McNamara} \& {Nulsen}(2007)}]{McNamara07}
{McNamara}, B.~R., \& {Nulsen}, P.~E.~J. 2007, \araa, 45, 117

\bibitem[{{McNamara} \& {O'Connell}(1989)}]{McNamara89}
{McNamara}, B.~R., \& {O'Connell}, R.~W. 1989, \aj, 98, 2018

\bibitem[{{Meece} {et~al.}(2015){Meece}, {O'Shea}, \& {Voit}}]{Meece15}
{Meece}, G.~R., {O'Shea}, B.~W., \& {Voit}, G.~M. 2015, \apj, 808, 43

\bibitem[{{Miniati}(2015)}]{Miniati15}
{Miniati}, F. 2015, \apj, 800, 60

\bibitem[{{Mittal} {et~al.}(2009){Mittal}, {Hudson}, {Reiprich}, \&
  {Clarke}}]{Mittal09}
{Mittal}, R., {Hudson}, D.~S., {Reiprich}, T.~H., \& {Clarke}, T. 2009, \aap,
  501, 835

\bibitem[{{Morsony} {et~al.}(2010){Morsony}, {Heinz}, {Br{\"u}ggen}, \&
  {Ruszkowski}}]{Morsony10}
{Morsony}, B.~J., {Heinz}, S., {Br{\"u}ggen}, M., \& {Ruszkowski}, M. 2010,
  \mnras, 407, 1277

\bibitem[{{Narayan} \& {Medvedev}(2001)}]{Narayan01}
{Narayan}, R., \& {Medvedev}, M.~V. 2001, \apjl, 562, L129

\bibitem[{{Navarro} {et~al.}(1996){Navarro}, {Frenk}, \& {White}}]{Navarro96}
{Navarro}, J.~F., {Frenk}, C.~S., \& {White}, S.~D.~M. 1996, \apj, 462, 563

\bibitem[{{Nulsen} {et~al.}(2005){Nulsen}, {McNamara}, {Wise}, \&
  {David}}]{Nulsen05}
{Nulsen}, P.~E.~J., {McNamara}, B.~R., {Wise}, M.~W., \& {David}, L.~P. 2005,
  \apj, 628, 629

\bibitem[{{O'Dea} {et~al.}(2008){O'Dea}, {Baum}, {Privon}, {et~al.}}]{ODea08}
{O'Dea}, C.~P., {Baum}, S.~A., {Privon}, G., {et~al.} 2008, \apj, 681, 1035

\bibitem[{{Panagoulia} {et~al.}(2014){Panagoulia}, {Fabian}, \&
  {Sanders}}]{Panagoulia14}
{Panagoulia}, E.~K., {Fabian}, A.~C., \& {Sanders}, J.~S. 2014, \mnras, 438,
  2341

\bibitem[{{Peterson} {et~al.}(2003){Peterson}, {Kahn}, {Paerels},
  {et~al.}}]{Peterson03}
{Peterson}, J.~R., {Kahn}, S.~M., {Paerels}, F.~B.~S., {et~al.} 2003, \apj,
  590, 207

\bibitem[{{Pfrommer}(2013)}]{Pfrommer13}
{Pfrommer}, C. 2013, \apj, 779, 10

\bibitem[{{Pfrommer} {et~al.}(2005){Pfrommer}, {En{\ss}lin}, \&
  {Sarazin}}]{Pfrommer05}
{Pfrommer}, C., {En{\ss}lin}, T.~A., \& {Sarazin}, C.~L. 2005, \aap, 430, 799

\bibitem[{{Pizzolato} \& {Soker}(2005)}]{Pizzolato05}
{Pizzolato}, F., \& {Soker}, N. 2005, \apj, 632, 821

\bibitem[{{Prasad} {et~al.}(2015){Prasad}, {Sharma}, \& {Babul}}]{Prasad15}
{Prasad}, D., {Sharma}, P., \& {Babul}, A. 2015, \apj, 811, 108

\bibitem[{{Rafferty} {et~al.}(2006){Rafferty}, {McNamara}, {Nulsen}, \&
  {Wise}}]{Rafferty06}
{Rafferty}, D.~A., {McNamara}, B.~R., {Nulsen}, P.~E.~J., \& {Wise}, M.~W.
  2006, \apj, 652, 216

\bibitem[{{Randall} {et~al.}(2015){Randall}, {Nulsen}, {Jones},
  {et~al.}}]{Randall15}
{Randall}, S.~W., {Nulsen}, P.~E.~J., {Jones}, C., {et~al.} 2015, \apj, 805,
  112

\bibitem[{{Reynolds} {et~al.}(2015){Reynolds}, {Balbus}, \&
  {Schekochihin}}]{Reynolds15}
{Reynolds}, C.~S., {Balbus}, S.~A., \& {Schekochihin}, A.~A. 2015, \apj, 815,
  41

\bibitem[{{Reynolds} {et~al.}(2005){Reynolds}, {McKernan}, {Fabian}, {Stone},
  \& {Vernaleo}}]{Reynolds05}
{Reynolds}, C.~S., {McKernan}, B., {Fabian}, A.~C., {Stone}, J.~M., \&
  {Vernaleo}, J.~C. 2005, \mnras, 357, 242

\bibitem[{{Robinson} {et~al.}(2004){Robinson}, {Dursi}, {Ricker},
  {et~al.}}]{Robinson04}
{Robinson}, K., {Dursi}, L.~J., {Ricker}, P.~M., {et~al.} 2004, \apj, 601, 621

\bibitem[{{Ruszkowski} {et~al.}(2004){Ruszkowski}, {Br{\"u}ggen}, \&
  {Begelman}}]{Ruszkowski04b}
{Ruszkowski}, M., {Br{\"u}ggen}, M., \& {Begelman}, M.~C. 2004, \apj, 615, 675

\bibitem[{{Ruszkowski} {et~al.}(2008){Ruszkowski}, {En{\ss}lin}, {Br{\"u}ggen},
  {Begelman}, \& {Churazov}}]{Ruszkowski08}
{Ruszkowski}, M., {En{\ss}lin}, T.~A., {Br{\"u}ggen}, M., {Begelman}, M.~C., \&
  {Churazov}, E. 2008, \mnras, 383, 1359

\bibitem[{{Ruszkowski} \& {Oh}(2011)}]{Ruszkowski11a}
{Ruszkowski}, M., \& {Oh}, S.~P. 2011, \mnras, 414, 1493

\bibitem[{{Ryu} {et~al.}(2003){Ryu}, {Kang}, {Hallman}, \& {Jones}}]{Ryu03}
{Ryu}, D., {Kang}, H., {Hallman}, E., \& {Jones}, T.~W. 2003, \apj, 593, 599

\bibitem[{{Salom{\'e}} \& {Combes}(2003)}]{Salome03}
{Salom{\'e}}, P., \& {Combes}, F. 2003, \aap, 412, 657

\bibitem[{{Sanders} {et~al.}(2010){Sanders}, {Fabian}, {Frank}, {Peterson}, \&
  {Russell}}]{Sanders10}
{Sanders}, J.~S., {Fabian}, A.~C., {Frank}, K.~A., {Peterson}, J.~R., \&
  {Russell}, H.~R. 2010, \mnras, 402, 127

\bibitem[{{Schekochihin} \& {Cowley}(2007)}]{Schekochihin07}
{Schekochihin}, A.~A., \& {Cowley}, S.~C. 2007, {Turbulence and Magnetic Fields
  in Astrophysical Plasmas}, ed. S.~{Molokov}, R.~{Moreau}, \& H.~K. {Moffatt}
  (Springer), 85

\bibitem[{{Sharma} {et~al.}(2012){Sharma}, {McCourt}, {Quataert}, \&
  {Parrish}}]{Sharma12}
{Sharma}, P., {McCourt}, M., {Quataert}, E., \& {Parrish}, I.~J. 2012, \mnras,
  420, 3174

\bibitem[{{Sharma} {et~al.}(2010){Sharma}, {Parrish}, \& {Quataert}}]{Sharma10}
{Sharma}, P., {Parrish}, I.~J., \& {Quataert}, E. 2010, \apj, 720, 652

\bibitem[{{Sijacki} {et~al.}(2007){Sijacki}, {Springel}, {Di Matteo}, \&
  {Hernquist}}]{Sijacki07}
{Sijacki}, D., {Springel}, V., {Di Matteo}, T., \& {Hernquist}, L. 2007,
  \mnras, 380, 877

\bibitem[{{Soker}(2003)}]{Soker03}
{Soker}, N. 2003, \mnras, 342, 463

\bibitem[{{Soker} {et~al.}(2015){Soker}, {Hillel}, \& {Sternberg}}]{Soker15}
{Soker}, N., {Hillel}, S., \& {Sternberg}, A. 2015, arXiv: 1504.07754

\bibitem[{{Stewart} {et~al.}(1984){Stewart}, {Fabian}, {Nulsen}, \&
  {Canizares}}]{Stewart84}
{Stewart}, G.~C., {Fabian}, A.~C., {Nulsen}, P.~E.~J., \& {Canizares}, C.~R.
  1984, \apj, 278, 536

\bibitem[{{Sutherland} \& {Dopita}(1993)}]{SutherlandDopita}
{Sutherland}, R.~S., \& {Dopita}, M.~A. 1993, \apjs, 88, 253

\bibitem[{{Takahashi} {et~al.}(2016){Takahashi}, {Kokubun}, {Mitsuda},
  {et~al.}}]{Takahashi16}
{Takahashi}, T., {Kokubun}, M., {Mitsuda}, K., {et~al.} 2016, Nature, submitted

\bibitem[{{Vazza} {et~al.}(2013){Vazza}, {Br{\"u}ggen}, \& {Gheller}}]{Vazza13}
{Vazza}, F., {Br{\"u}ggen}, M., \& {Gheller}, C. 2013, \mnras, 428, 2366

\bibitem[{{Vernaleo} \& {Reynolds}(2006)}]{Vernaleo06}
{Vernaleo}, J.~C., \& {Reynolds}, C.~S. 2006, \apj, 645, 83

\bibitem[{{Voigt} \& {Fabian}(2004)}]{Voigt04}
{Voigt}, L.~M., \& {Fabian}, A.~C. 2004, \mnras, 347, 1130

\bibitem[{{Voit} {et~al.}(2015){Voit}, {Donahue}, {Bryan}, \&
  {McDonald}}]{Voit15}
{Voit}, G.~M., {Donahue}, M., {Bryan}, G.~L., \& {McDonald}, M. 2015, \nat,
  519, 203

\bibitem[{{Yang} \& {Reynolds}(2016)}]{Yang15}
{Yang}, H.-Y.~K., \& {Reynolds}, C.~S. 2016, \apj, 818, 181

\bibitem[{{Yang} {et~al.}(2012{\natexlab{a}}){Yang}, {Ruszkowski}, {Ricker},
  {Zweibel}, \& {Lee}}]{Yang12}
{Yang}, H.-Y.~K., {Ruszkowski}, M., {Ricker}, P.~M., {Zweibel}, E., \& {Lee},
  D. 2012{\natexlab{a}}, \apj, 761, 185

\bibitem[{{Yang} {et~al.}(2012{\natexlab{b}}){Yang}, {Sutter}, \&
  {Ricker}}]{Yang12b}
{Yang}, H.-Y.~K., {Sutter}, P.~M., \& {Ricker}, P.~M. 2012{\natexlab{b}},
  \mnras, 427, 1614

\bibitem[{{Zakamska} \& {Narayan}(2003)}]{Zakamska03}
{Zakamska}, N.~L., \& {Narayan}, R. 2003, \apj, 582, 162

\bibitem[{{Zhdankin} {et~al.}(2015){Zhdankin}, {Uzdensky}, \&
  {Boldyrev}}]{Zhdankin15}
{Zhdankin}, V., {Uzdensky}, D.~A., \& {Boldyrev}, S. 2015, \apj, 811, 6

\bibitem[{{Zhuravleva} {et~al.}(2016){Zhuravleva}, {Churazov}, {Ar{\'e}valo},
  {et~al.}}]{Zhuravleva16}
{Zhuravleva}, I., {Churazov}, E., {Ar{\'e}valo}, P., {et~al.} 2016, \mnras,
  458, 2902

\bibitem[{{Zhuravleva} {et~al.}(2014){Zhuravleva}, {Churazov}, {Schekochihin},
  {et~al.}}]{Zhuravleva14}
{Zhuravleva}, I., {Churazov}, E., {Schekochihin}, A.~A., {et~al.} 2014, \nat,
  515, 85

\bibitem[{{ZuHone} {et~al.}(2013){ZuHone}, {Markevitch}, {Brunetti}, \&
  {Giacintucci}}]{Zuhone13}
{ZuHone}, J.~A., {Markevitch}, M., {Brunetti}, G., \& {Giacintucci}, S. 2013,
  \apj, 762, 78

\end{thebibliography}

\end{document}